\documentclass[11pt]{article}
\usepackage{graphicx,tabularx,epsfig}
\usepackage{color}
\usepackage{enumitem}
\usepackage{caption}
\usepackage{subcaption}
\usepackage{amsmath}
\usepackage{amssymb}
\usepackage{amsthm}
\usepackage{physics}
\usepackage{dsfont, mathtools}
\usepackage{psvectorian}
\usepackage{blkarray}
\usepackage{bm}
\usepackage{url}
\usepackage{setspace}

\usepackage{xcolor}
\colorlet{linkequation}{blue}

\usepackage[colorlinks = true,
            linkcolor = blue,
            urlcolor  = blue,
            citecolor = blue,
            anchorcolor = blue]{hyperref}

\usepackage{floatrow}

\newlength{\abstractwidth}
\renewcommand{\thefootnote}{\fnsymbol{footnote}}
\renewcommand{\thanks}[1]{\footnote{#1}} 
\newcommand{\starttext}{
\setcounter{footnote}{0}
\renewcommand{\thefootnote}{\arabic{footnote}}}

\makeatletter
\g@addto@macro\normalsize{%
  \setlength\abovedisplayskip{15pt}
  \setlength\belowdisplayskip{15pt}
  \setlength\abovedisplayshortskip{15pt}
  \setlength\belowdisplayshortskip{15pt}
}
\makeatother

\usepackage{color}

\renewcommand{\title}[1]{\vbox{\center\LARGE{#1}}\vspace{5mm}}
\renewcommand{\author}[1]{\vbox{\center#1}\vspace{5mm}}

\begin{document}

\singlespacing

\begin{center}

{\Large \bf {Operator growth and black hole formation}}

\bigskip \noindent

\bigskip

 {\bf Felix M.\ Haehl$^a$ and Ying Zhao$^b$}

    \vspace{0.5in}

    $^a$School of Mathematical Sciences, University of Southampton, SO17 1BJ, U.K. \vskip1em
    $^b$Kavli Institute for Theoretical Physics, Santa Barbara, CA 93106, USA
    
    \vspace{0.5in}
    
    {\tt f.m.haehl@soton.ac.uk, zhaoying@kitp.ucsb.edu}

\bigskip

\end{center}

\begin{abstract}

When two particles collide in an asymptotically AdS spacetime with high enough energy and small enough impact parameter, they can form a black hole. Motivated by dual quantum circuit considerations, we propose a threshold condition for black hole formation. Intuitively the condition can be understood as the onset of overlap of the butterfly cones describing the ballistic spread of the effect of the perturbations on the boundary systems. We verify the correctness of the condition in three bulk dimensions. We describe a six-point correlation function that can diagnose this condition and compute it in two-dimensional CFTs using eikonal resummation.

\medskip
\noindent
\end{abstract}

\newpage

\starttext \baselineskip=17.63pt \setcounter{footnote}{0}

{\hypersetup{hidelinks}
\tableofcontents
}

\newpage
\section{Introduction}
\label{sec:intro}

Despite extensive research within the framework of AdS/CFT duality \cite{Maldacena:1997re,Gubser:1998bc,Witten:1998qj}, black holes remain puzzling objects. More so, their formation is not properly understood. The spacetime shows a very different behavior before and after a black hole forms: prior to black hole formation the spacetime is static. After a black hole forms, the part of spacetime in the interior rapidly grows, and a singularity emerges where the classical description of spacetime fails. 

\begin{figure}[b] 
 \begin{center}                      
      \includegraphics[width=3in]{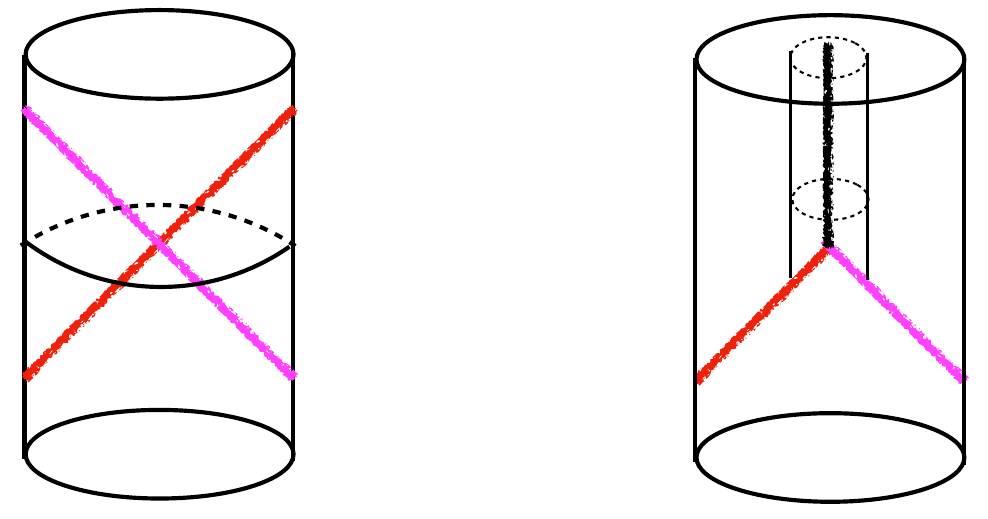}
      \caption{Cartoon of the collision of two localized particles sent into an AdS-Rindler spacetime from the boundary. At high enough energies and small enough impact parameter, a black hole forms.}
  \label{collision_AdS}
  \end{center}
\end{figure}

It would be desirable to have a microscopic understanding of black hole formation, which might teach us new lessons about black holes as well as quantum gravity in general. In this paper we give a new perspective on this problem (see \cite{Balasubramanian:2000rt,Anous:2016kss,Anous:2019yku} for related work). We consider sending two particles into the AdS spacetime from near the boundary. When the two particles collide with high enough energy or small enough impact parameter, they can form a black hole, see figure \ref{collision_AdS}.  We will give a threshold condition for black hole formation in terms of a particular microscopic description of the collision process.

There are various coordinates describing the AdS spacetime. For our purpose, we will use AdS-Rindler coordinates, as these respect various symmetries of the collision process, in particular translation symmetry in the transverse direction, as well as boost symmetry in the time direction. AdS-Rindler space is very similar to an eternal black hole; more precisely, it is an eternal hyperbolic black hole with temperature $\frac{1}{2\pi \ell_{AdS}}$. If we consider the collision as happening inside the interior of an eternal hyperbolic black hole (figure \ref{butterfly_overlap}), then there exists a quantum information theoretic boundary description of this collision process: its basic features map to properties of two perturbations spreading exponentially and ballistically in a shared quantum circuit. This picture was proposed in \cite{Zhao:2020gxq} and tested in detail for different setups and kinematic regimes in \cite{Haehl:2021prg,Haehl:2021tft,Haehl:2022frr}.

Using this approach, we propose a threshold condition for the formation of a black hole. Each perturbation is characterized by a causal cone and a butterfly cone, describing respectively its causal impact on other degrees of freedom as well as the disruption of the entanglement structure of the state as the effect of the operator spreads throughout the system ballistically \cite{Roberts:2014isa}. We propose that the collision creates a black hole in $AdS$ precisely when the two butterfly cones have overlap in the shared quantum circuit (figure \ref{butterfly_overlap}).

The geometry of black hole formation from particle collisions has been worked out in $AdS_3$ \cite{Gott:1990zr,Matschull:1998rv,Holst:1999tc}. We confirm that our threshold condition is consistent with these results. We expect our condition to work in higher dimensions we well, when the impact parameter is large compared to the AdS scale. 

Looking ahead, the most interesting question to ask is what such a connection with quantum circuit models teaches us about black hole formation. The dynamics shows completely different features depending on whether the butterfly cones do or do not overlap. This suggests a deep connection between the formation of black holes in AdS and the dynamics of quantum chaos in the boundary description.

This paper is organized as follows. In section \ref{sec:review} we review operator growth and the quantum circuit picture of the particle collision. In section \ref{sec:threshold} we give the threshold condition for black hole formation and motivate it from both the point of view of quantum circuits as well as gravity. In section \ref{sec:diagnosis} we propose a CFT six-point function that can be used to diagnose this threshold condition. In section \ref{sec:calculations} we compute the `gravitational' contribution to this six-point function in two-dimensional CFTs explicitly, using recently developed eikonal methods as well as the method of boundary reparametrizations. We point out further questions and future directions in section \ref{sec:discussion}.

\section{Review: operator growth and quantum circuits}
\label{sec:review}

In \cite{Zhao:2020gxq,Haehl:2022frr} we studied particle collisions on an eternal black hole background. We gave a quantum circuit interpretation of the collision process, and studied the properties of spacetime geometry after the collision. In this section we will study particle collisions in $AdS$ by treating them as collisions on an eternal hyperbolic black hole background. Our focus will be the regime of strong collisions.

\subsection{Causal cone and butterfly cone}

We begin with a brief review of the quantum circuit picture for one particle propagating on a black hole background (see \cite{Stanford:2014jda, Susskind:2014jwa,Roberts:2014isa,Susskind:2018tei} for more details). 

In \cite{Susskind:2018tei} it was pointed out that the process of a particle falling into a black hole is closely related to the growth of an operator in the dual boundary field theory. Consider a holographic field theory at finite temperature whose bulk dual contains a black hole. We apply a local operator to the boundary at time $t_1$ and location $x_1$. The effect of the operator spreads ballistically into transverse directions. At the same time, the operator also grows within the matrix degrees of freedom. As the interactions among the matrix degrees of freedom is all-to-all, this growth is exponential in time, with Lyapunov exponent given by $\frac{2\pi}{\beta}$.

To get an intuitive picture, one can consider a model of a chain of SYK models \cite{Gu:2016oyy}. The Hamiltonian of the SYK chain contains the sum of Hamiltonians for each SYK `dot' plus some simple couplings between neighboring sites. Note that within each copy of the SYK system, the interaction is $k$-local, i.e., every fermion is coupled to every other fermion within the same SYK but each term in Hamiltonian contains at most $k$ fermions. On the other hand, different SYK sites only interact with their nearest neighbors. We act with a simple operator on one of the SYK systems along the chain. This perturbation will grow exponentially on its site with a Lyapunov exponent depending on the temperature. At the same time, the perturbation will also grow to other SYK nodes on the chain in a ballistic fashion.
\begin{figure}
 \begin{center}                      
      \includegraphics[width=2.2in]{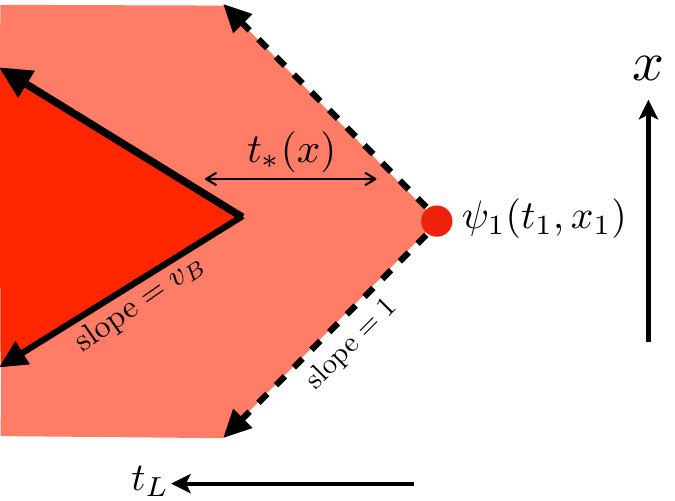}
      \caption{Illustration of the causal cone (light red) and butterfly cone (dark red) describing the relativistic spread in space and the ballistic exponential growth of a small perturbation.}
  \label{two_cones}
  \end{center}
\end{figure}
A similar picture is expected to apply to a holographic conformal field theory. 
Following \cite{Roberts:2014isa}, one can summarize these two notions of growth of a local perturbation in terms of a causal and a butterfly cone (in spacetime).\footnote{Note that the SYK chain does not actually have a causal cone, as it is not a relativistic field theory. The two-cone structure does however exist in holographic CFTs and we shall argue in this paper that quantities motivated in the SYK model indeed serve as good diagnostics for black hole formation in higher dimensional field theories.} In figure \ref{two_cones}, the vertical direction represents space while the horizontal direction represents time. We apply a simple operator at the tip of the causal cone (pale red) outside of which the perturbation cannot have any effect. The dark red region is the butterfly cone, which delineates where the perturbation not only has {\it any} effect, but its effect becomes large and affects most of the local (matrix) degrees of freedom. In the simplified example of the SYK chain, all fermions in a given SYK node are affected when the node enters the butterfly cone. For the butterfly cone, we have $\frac{dx}{dt}\equiv v_B\leq 1$. The horizontal time difference between the edges of the two cones is the scrambling time $t_*$, i.e., the time it takes for the perturbation from entering a local set of degrees of freedom to affecting all of them.

\subsection{Motivation: operator size and out-of-time-ordered correlators}
\label{sec:motivation}

We quantify the process of operator growth by the {\it size} of an operator \cite{Roberts:2018mnp,Qi:2018bje}. We again start from the SYK toy model, where this can be made precise. The infinite temperature size $n_\infty$ of an operator is defined as the average number of fermions contained in the operator (thought of as a string of fermions) \cite{Roberts:2018mnp}. This counting problem turns out to be computable by:
\begin{align}
	n_{\infty}[\mathcal{O}] \equiv \frac{1}{4}\sum_j\tr(\{\mathcal{O},\psi_j\}^{\dagger}\{\mathcal{O},\psi_j\}) \,.
\end{align}
At finite temperature $\frac{1}{\beta}$, we consider a background state $\rho = e^{-\beta H}$ and use the following renormalized and regularized definition of operator size \cite{Qi:2018bje}:
\begin{align}
\label{size_1}
	\frac{n_{\beta}[\mathcal{O}]}{n_{\text{max}}}\equiv \frac{n_{\infty}[\mathcal{O}\rho^{\frac{1}{2}}]-n_{\infty}[\rho^{\frac{1}{2}}]}{n_{\text{max}}-n_{\infty}[\rho^{\frac{1}{2}}]}\,,\qquad\quad n_\text{max} \equiv \frac{N}{2} \,.
\end{align}
Note that \eqref{size_1} is related to an out-of-time-ordered four-point function. It is small for early times and then grows exponentially until saturation to $n_{max}$ at scrambling time $t_*$. For instance:
\begin{align}
	\mathcal{F}_4(t_1)\equiv 1-\frac{n_{\beta}[\psi(t_1)]}{n_{\text{max}}} = -\frac{\sum_{j = 1}^N\expval{\psi_j(i\frac{\beta}{2})\psi_1(t_1)\psi_j\psi_1(t_1)}_{\beta}}{\sum_{j = 1}^N\expval{\psi_1(t_1)\psi_1(t_1)}_{\beta}\expval{\psi_j(i\frac{\beta}{2})\psi_j}_{\beta}}  \,.
\end{align}
Next, we consider SYK chain and perturb it at time $t_1$ and location $x_1$. To measure the effect of the perturbation at location $x$ (at time $t=0$), we again consider the same four-point function:
\begin{align}
\label{cone_F4}
	\mathcal{F}_4(x)\equiv 1-\frac{n_{\beta}[\psi_1(t_1,x_1),x]}{n_{\text{max}}} =  -\frac{\sum_{j = 1}^N\expval{\psi_j(i\frac{\beta}{2},x)\psi_1(t_1,x_1)\psi_j(x)\psi_1(t_1,x_1)}_{\beta}}{\sum_{j = 1}^N\expval{\psi_1(t_1,x_1)\psi_1(t_1,x_1)}_{\beta}\expval{\psi_j(i\frac{\beta}{2},x)\psi_j(x)}_{\beta}}
\end{align}

A general definition of operator size in a holographic field theories so far does not exist.\footnote{There have been different interesting attempts, such as direct generalizations of the notion of operator size we use here \cite{Magan:2020iac,Gu:2021xaj,Haehl:2021emt}, as well as connections with operator complexity \cite{Magan:2018nmu,Parker:2018yvk,Barbon:2019wsy,Rabinovici:2020ryf,Kar:2021nbm} and related work on the gravity picture \cite{Jafferis:2020ora,Leutheusser:2021qhd,Nomura:2018kia}.} Motivated by the SYK chain model, we will use the four-point function as in \eqref{cone_F4} to characterize the growth of an operator. One feature to notice in \eqref{cone_F4} is the sharp transition across the boundary of the butterfly cone.

\subsection{Overlap of causal cones and particle collision}

\begin{figure}
 \begin{center}                      
      \includegraphics[width=2.2
      in]{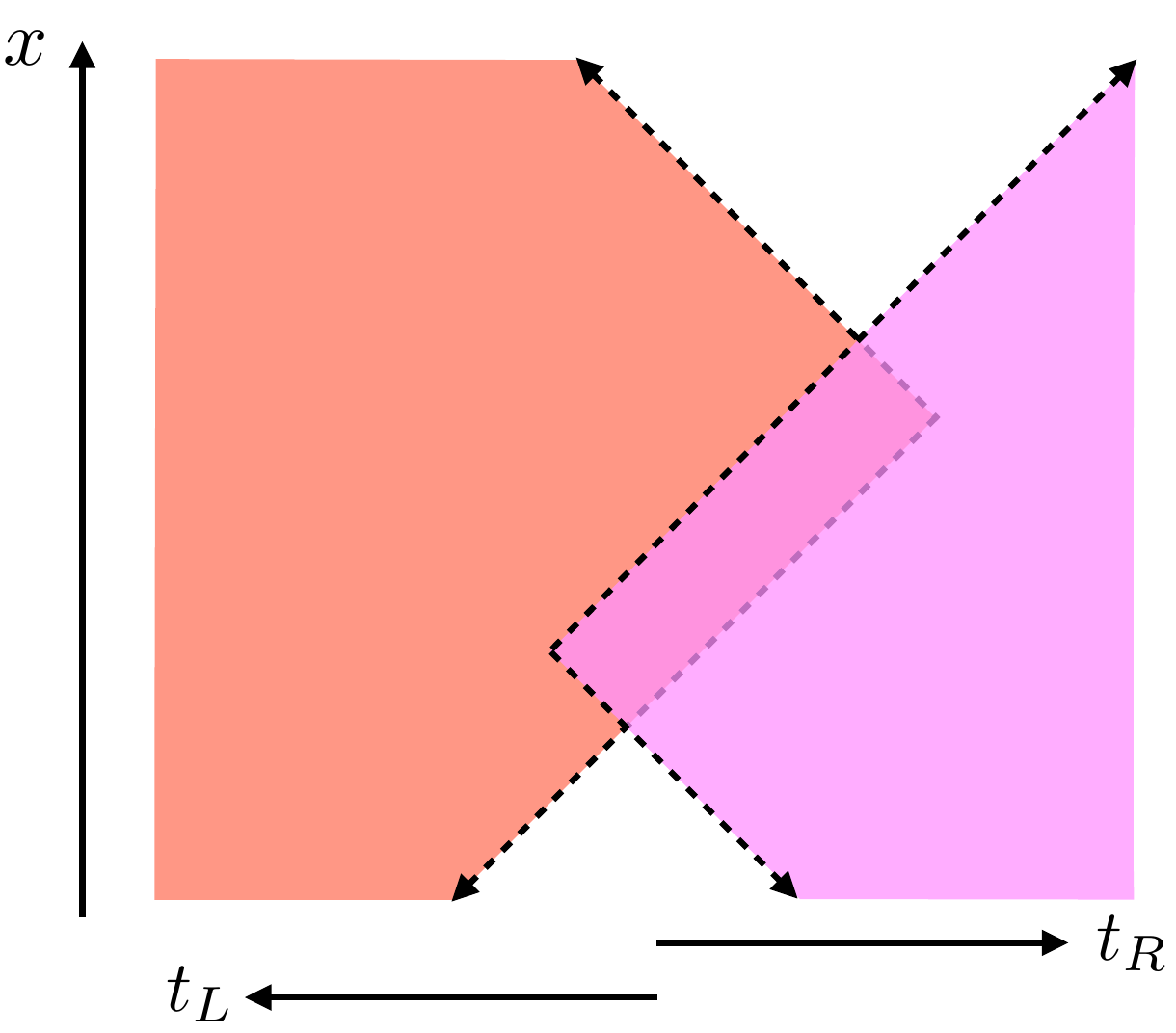}
      \caption{Overlap of two causal cones spreading in opposite directions of the quantum circuit (i.e., one of them corresponds to a perturbation of the right system, and the other to a perturbation of the left system). Each constant $x$ slice can be thought of as a local quantum circuit describing the evolution of the thermofield double state shared between local matrix degrees of freedom in the two systems at location $x$.}
  \label{causal_cone_overlap}
  \end{center}
\end{figure}

Now consider two systems entangled in themofield double state. Such a setup enjoys a dual description in terms of the eternal Schwarschild black hole \cite{Maldacena:2001kr}. Applying a local operator to each of the systems corresponds to two particles falling into the black hole from each of the boundaries, which we can approximate as localized gravitational shocks. The shared state of the two systems and its evolution can be described in terms of a quantum circuit consisting of a collection of qubits on which unitary operations are performed. This provides a dual quantum mechanical description of the bulk region inside the wormhole \cite{Hartman:2013qma,Susskind:2014moa}. Each of the perturbations spreads within this quantum circuit. In \cite{Zhao:2020gxq,Haehl:2022frr} it was argued that the two particles gravitationally interact with each other in the bulk if and only if the two causal cones overlap in the shared quantum circuit (figure \ref{causal_cone_overlap}). 
In \cite{Zhao:2020gxq,Haehl:2022frr} we studied the post-collision geometry and compared it to properties of the quantum circuit. For collisions of localized shocks, the post-collision geometry is generally unknown. In \cite{Haehl:2022frr} we used approximations that are valid in a regime when the collision is not too strong. In this paper our focus will be on the complementary regime when the collision is indeed strong enough to form a black hole.

\section{Threshold for black hole formation}
\label{sec:threshold}

Starting from vacuum AdS, we send in two particles which scatter in the bulk. The collision may or may not be head on. We can describe this process in Rindler coordinates and put each of the particles into one of the complementary Rindler wedges (figure \ref{collision_AdS}, left panel). We will arrange the times such that the collision happens close to the Rindler horizon. 

 \begin{figure}
 \begin{center}                      
      \includegraphics[width=.9\textwidth]{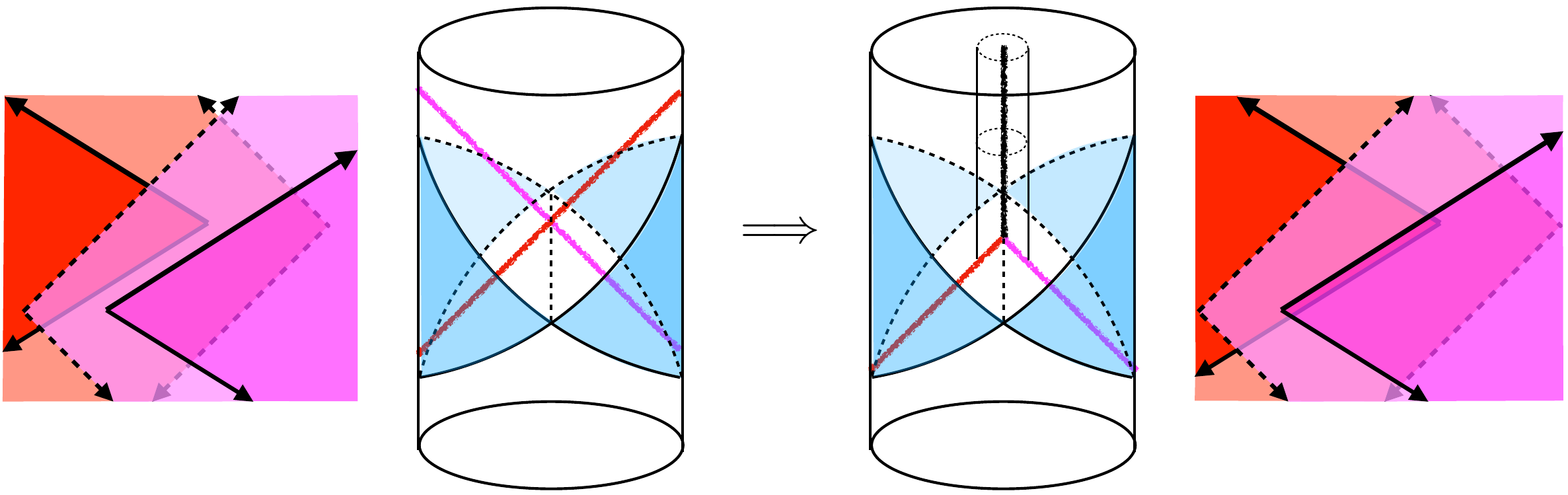}
      \caption{Collision setup in Rindler AdS and in the quantum circuit. We propose that a black hole forms precisely when the butterfly cones begin to overlap (right hand side).}
  \label{butterfly_overlap}
  \end{center}
\end{figure}

Our proposal is as follows: {\it for such a collision process, black hole formation occurs when the two butterfly cones in the corresponding quantum circuit overlap.} Such a configuration is illustrated on the right of figure \ref{butterfly_overlap}. We will now present motivation and evidence for this proposal.

\subsection{Motivation from quantum circuits}

\begin{figure}
 \begin{center}                     
      \includegraphics[width=\textwidth]{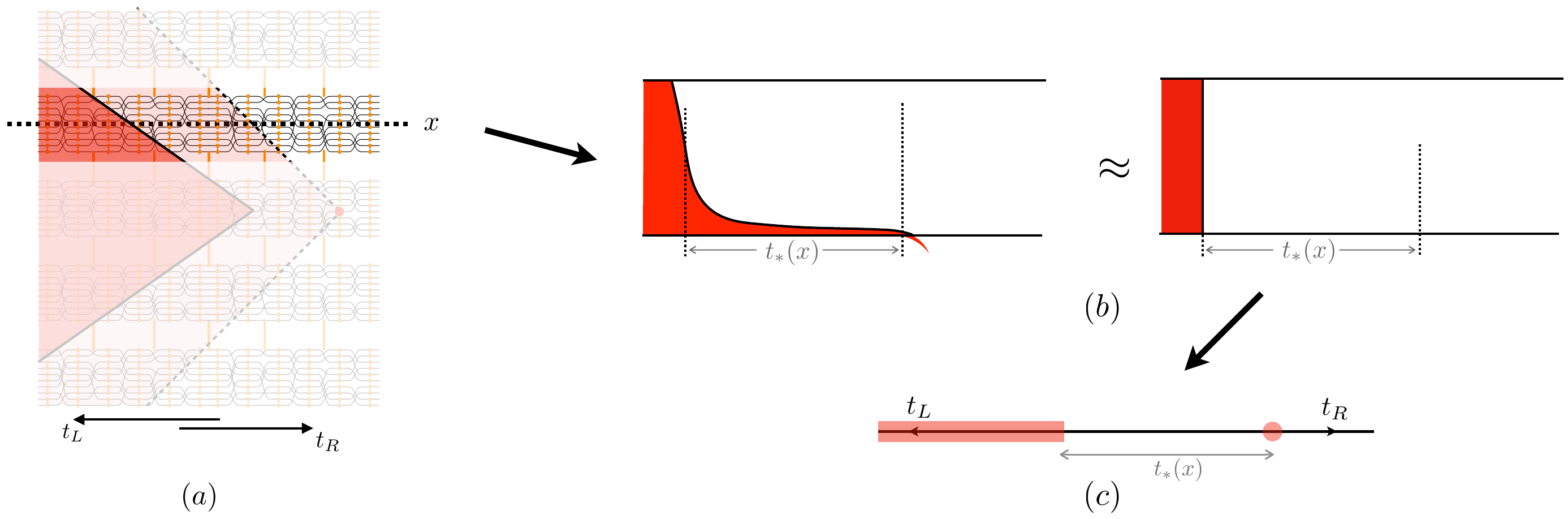}
      \caption{(a) Each transverse $x$-slice has a description in terms of a quantum circuit with all-to-all qubit interactions. (b) Exponential growth of the perturbation within this circuit can be approximated by a step function located at a time that is separated from the initial perturbation by $t_*(x)$. (c) We can further simplify notation and collapse the approximate diagram into a line and denote the intersection of the $x$-slice with the causal cone of the perturbation by a faint dot; we will use this notation in section \ref{sec:diagnosis}.}
  \label{approximation_x1}
  \end{center}
\end{figure}

Consider the circuit picture and focus on some particular transverse location $x$ (figure \ref{approximation_x1}). Each perturbation spreads exponentially through the local circuit for $x$ once the causal cone reaches this location.
We use pictures such as figure \ref{approximation_x1}(b) to represent the exponential growth through the local quantum circuit at a fixed transverse location $x$ (the vertical direction is then not spatial, but rather represents the stack of qubits, i.e., the width of the circuit). To first approximation, the growth of the perturbation at $x$ corresponds to a step function because the exponential growth has a small effect until a scrambling time $t_*(x)$ later, where it rapidly affects the entire system.

\begin{figure} 
 \begin{center}                      
      \includegraphics[width=4.5
      in]{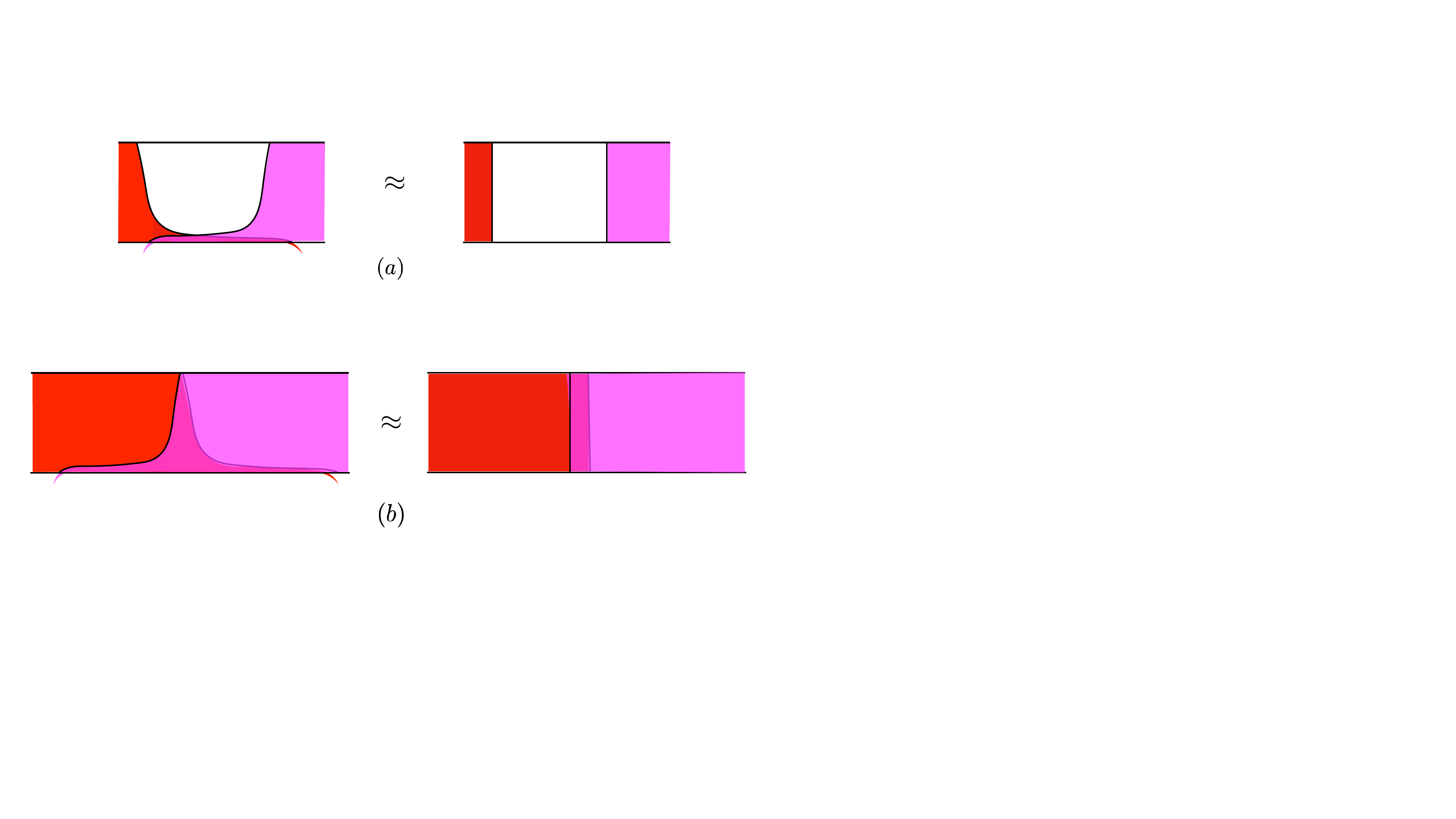}
      \caption{There is a sharp transition in the quantum circuit when the two butterfly cones overlap. The step function approximation describes constant $x$ slices of the circuit pictures in figure \ref{butterfly_overlap}.}
  \label{transition_sharp}
  \end{center}
\end{figure}

Now we consider the overlap of two such perturbations. Figure \ref{transition_sharp}(a) shows the circuit at fixed transverse location in a configuration where the two causal cones overlap but the two butterfly cones do not. In figure \ref{transition_sharp}(b), the two butterfly cones overlap as well. In the step function approximation, we see that there is a sharp transition when the two butterfly cones start to overlap. We will quantify this sharp transition later. 

Two butterfly cones overlapping at transverse location $x$ is exactly the same statement that at circuit location $x$ there are no more quantum gates that are unaffected by both perturbations. As we see a sharp transition in the quantum circuit model, one expects a dramatic effect on the gravity side. In the case of particles scattering in $AdS$, it is a natural guess that this dramatic transition corresponds to black hole formation (see figure \ref{butterfly_overlap}).

\subsection{Motivation from gravity}

In this section we argue that our proposal is reasonable from purely gravitational considerations.

\begin{figure}[h]
 \begin{center}                      
      \includegraphics[width=3.6
      in]{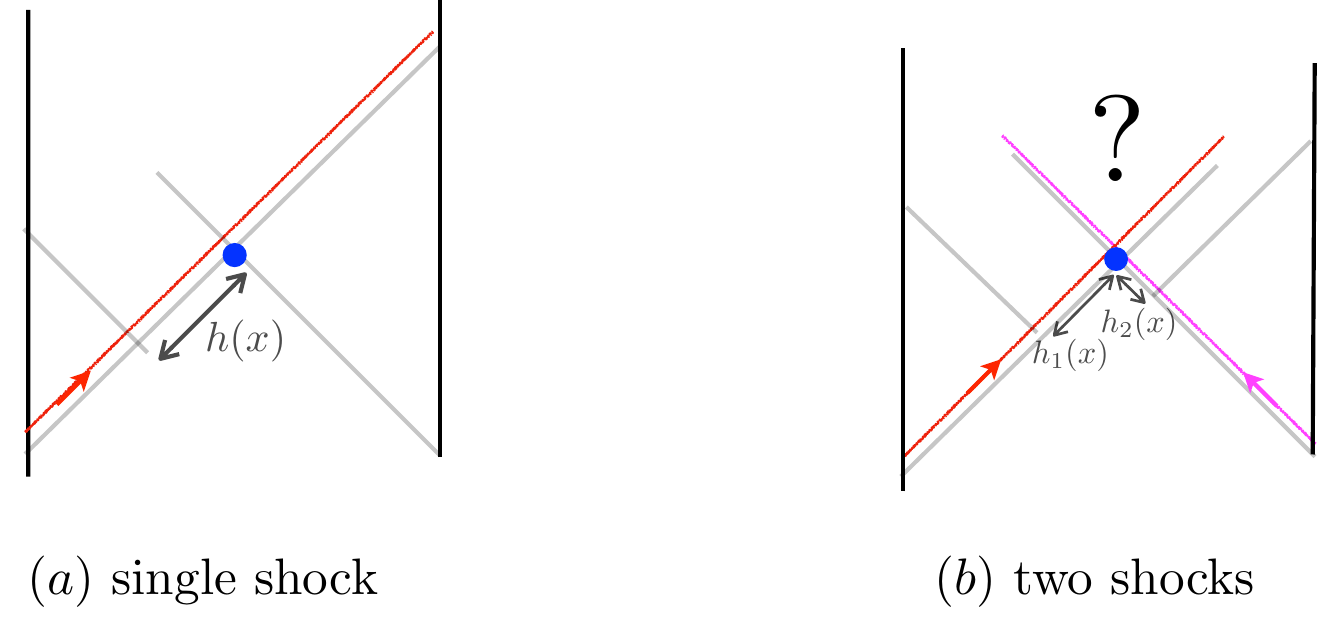}
      \caption{Geometry outside the post-collision region. The Kruskal coordinate shifts $h_{1,2}(x)$ characterize the strength of the shocks.}
  \label{outside}
  \end{center}
\end{figure}

Outside of the region to the causal future of the collision (`post-collision region'), the gravity solution is known. The corresponding Kruskal coordinate shifts along the shockwave by an amount $h(x)$ (figure \ref{outside}). Roughly speaking, the amount of shift decreases exponentially in the distance to the source, with exponent $\frac{2\pi}{\beta v_B}$. This encodes the fact that the butterfly cone has slope $\frac{dx}{dt}  = v_B$. 
Far away from the source, the expression for $h(x)$ is simple. Now consider one perturbation coming in from the left at time $t_1$ and location $x_1$, another perturbation coming in from the right at time $t_2$ and location $x_2$. Let $h_1(x)$ be the shift due to the left perturbation, and $h_2(x)$ be the shift due the the right perturbation. One finds \cite{Roberts:2014isa}
\begin{equation}
\label{h}
\begin{aligned}
	h_1(x)\approx e^{-t_{1}-t_{*1}-\frac{|x-x_{1}|}{v_B}}
	\,,\qquad h_2(x)\approx e^{-t_{2}-t_{*2}-\frac{|x-x_{2}|}{v_B}}\,.
\end{aligned}
\end{equation}
where $t_{*1}$ and $t_{*2}$ are the two scrambling times, which are functions of the entropies of the perturbations. The quantities $h_{1,2}(x)$ encode the strength of the shocks.
On the other hand, the butterfly cones overlap if the following condition is satisfied (this is trivial to see in the quantum circuit diagrams):
\begin{align}
\label{condition}
\text{butterfly cone overlap} \qquad \Leftrightarrow \qquad 
	\frac{b}{v_B}\lessapprox -t_{1}-t_{2}-t_{*1}-t_{*2}+\mathcal{O}(\ell_{AdS}) \,,
\end{align}
where $b$ is the impact parameter. The reason we use `$\lessapprox$' is related to the fact that the location of butterfly cone is defined up to thermal time, which is $\mathcal{O}(\ell_{AdS})$ in our case.

Comparing \eqref{condition} and \eqref{h}, one can easily see that the two butterfly cones overlapping is equivalent to the condition that $\max_x h_1(x)h_2(x)\gtrapprox 1$. 

In \cite{Haehl:2022frr} we used an ansatz solution to study the post-collision geometry. That solution is a good approximation as long as $h_1(x)h_2(x)\ll 1$.  When $h_1(x)h_2(x)$ becomes of order one, non-linear effects in general relativity become important and one expects significant change. So it should not be surprising that this corresponds to the threshold of black hole formation.

\subsection{Black hole formation in $AdS_3$: Gott condition}

In three bulk dimensions the geometric solution of the black hole formation problem via particle collision is known \cite{Matschull:1998rv,Holst:1999tc}. In particular, we know the exact threshold for black hole formation. In flat space scattering, this threshold is often referred to as Gott condition \cite{Gott:1990zr}. 
In a collision with impact parameter $b$, it reads as follows:
\begin{align}
\label{condition_2}
 \text{Gott condition for black hole formation:} \qquad 	8 G_N E_{CM} > e^{\frac{b}{2}}\,,
\end{align}
where $E_{CM}$ is energy of each of the particles in the center of mass frame. 

For a head-on collision ($b=0$) this simplifies to $8 G_N E_{CM}>1$. In AdS${}_3$, this condition precisely captures the threshold for a BTZ black hole to form. One corollary of this condition is that the minimal size black hole we can form in $AdS_3$ has energy of order Planck mass. This is consistent with the black hole threshold in the spectrum of two-dimensional CFTs.

To complete the consistency check, we note that one can equivalently write \eqref{condition_2} as $b<-t_{1}-t_{2}-t_{*1}-t_{*2}+\mathcal{O}(\ell_{AdS})$. This is also the condition for the butterfly cones overlapping in a two-dimensional CFT (with $v_B=1$, which is the correct value in two dimensions).

\section{Diagnosing the threshold for black hole formation}
\label{sec:diagnosis}

So far we have argued that the threshold for black hole formation coincides with the threshold for butterfly cone overlap. 
In this section we refine this idea and establish a boundary CFT correlation function that can diagnose the formation of the black hole. 

Again, we focus on a particular transverse location $x$ in the quantum circuit. Let us understand better what it means for the butterfly cones to overlap at $x$. Here, we will use the graphical notation introduced in figure \ref{approximation_x1}(c) where the quantum circuit at $x$ is collapsed into a line indicating the directions of left and right time evolution. 

 \begin{figure}
 \begin{center}                      
      \includegraphics[width=4.5
      in]{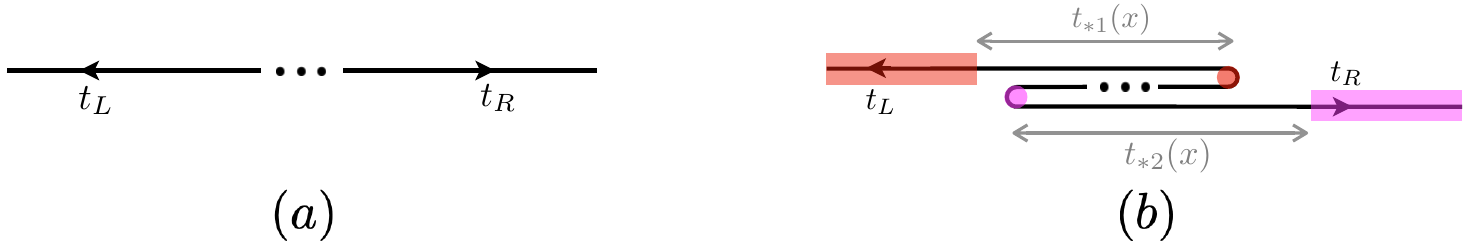}
      \caption{Quantum circuit at transverse location $x$. Intersections with the causal cones of the perturbations are shown as dots. The overlap of the two causal cones at $x$ manifests as a fold. The butterfly cones in this example do not overlap.}
  \label{circuit_x0}
  \end{center}
\end{figure}
Figure \ref{circuit_x0}(a) represents the quantum circuit at $x$ without any perturbation. The three dots in the middle indicate the entangled initial state. Figure \ref{circuit_x0}(b) shows an $x$-slice of the quantum circuit with two perturbations whose causal cones overlap at $x$. As in figure \ref{approximation_x1}(c), the colored dots represent the boundary of the causal cones at $x$, and the shaded regions represent the butterfly cones. In figure \ref{circuit_x0}(b), the two butterfly cones do not overlap at $x$. 
When the two butterfly cones do not overlap, the fold almost cancels, and the perturbations have no large effect on the circuit: 
\begin{equation}
 \label{eq:cancel}
                 \includegraphics[width=.67\textwidth]{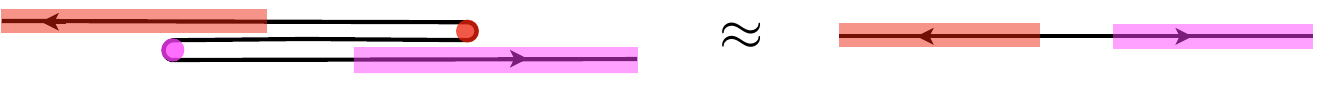}
\end{equation}
On the other hand, when the two butterfly cones do overlap at $x$, the fold in the circuit cannot cancel. The circuit will be significantly affected by the perturbations:
 \begin{equation}
 \label{eq:no_cancel}
                 \includegraphics[width=.75\textwidth]{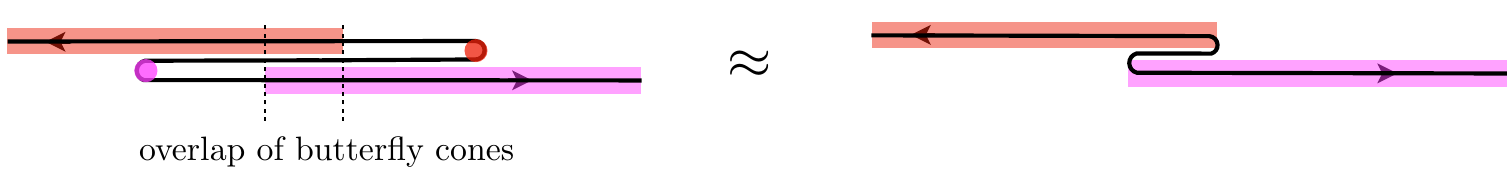}
\end{equation}

More quantitively, we can consider the size (as defined in section \ref{sec:motivation}) of the two-sided operator $e^{iHa}\psi_1(-t_{1},x_{1})\rho^{\frac{1}{2}}\psi_2(t_{2},x_{2})e^{iHb}$ at location $x$. In the SYK chain toy model, this is defined through the following six-point function: 
\begin{equation}
\label{6_point_function}
\begin{split}
	\mathcal{F}_6(x) &=1-\frac{n_\beta\big[e^{iHa} \psi_1 \rho^{\frac{1}{2}}\psi_2 e^{iHb},x\big]}{n_\text{max}} \\
	&\equiv\frac{\sum_{j = 1}^N\expval{\psi_1^L(t_{1},x_{1})\psi_2^R(t_{2},x_{2})\ \psi_j^L(a,x)\psi_j^R(b,x)\ \psi_1^L(t_{1},x_{1})\psi_2^R(t_{2},x_{2})}}{\sum_{j = 1}^N\expval{\psi_j^L(a,x)\psi_j^R(b,x)}}\,,
\end{split}
\end{equation}
where expectation values are taken in the thermofield double state.
To understand the significance of $\eqref{6_point_function}$ intuitively, one can represent it pictorially as\footnote{ By the size of a quantum circuit diagram, we mean the size of the appropriate state associated with it. More precisely, we mean \eqref{6_point_function}.}
\begin{equation}                    
      \includegraphics[width=5.6
      in]{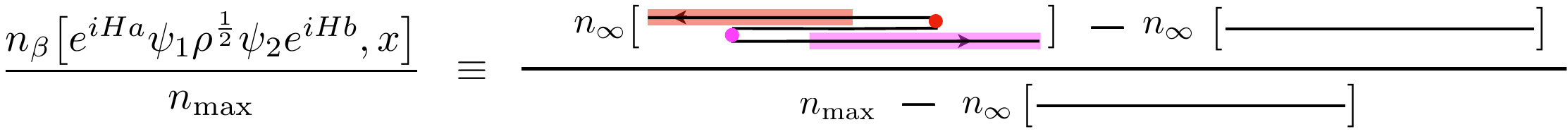}
       \label{size_def}
 \end{equation}
Comparing expressions \eqref{size_def} and \eqref{6_point_function}, and simplifying using \eqref{eq:cancel} and \eqref{eq:no_cancel}, one can easily see that for $x$ in between $x_{1}$ and $x_{2}$, the six-point function in \eqref{6_point_function} must be approximately $1$ when the butterfly cones do not overlap.  On the other hand, ${\cal F}_6(x)$ drops to zero when the butterfly cones do overlap and the corresponding collision creates a black hole. 

We wish to emphasize that the notion of operator size we used (which is derived for the SYK model) merely serves as a motivation for the particular six-point function we consider. Our arguments above demonstrate the use of this correlator as a diagnostic of black hole formation more generally. To strengthen this point, we will confirm the intuition provided here by computing the six-point function \eqref{6_point_function} explicitly in two-dimensional CFTs.

\section{Calculation in 2d CFT}
\label{sec:calculations}

Instead of the SYK six-point function \eqref{6_point_function}, we now consider a two-dimensional CFT and compute the relevant contribution to the following analogous quantity:
\begin{equation}
\label{eq:F6def2}
  {\cal F}_6 \overline{\cal F}_6 = \frac{\text{Tr} \left\{W_1(-t_1-i\delta_1,x_{1}) {\cal O}_j(-a,x_j) W_1 (-t_1,x_{1}) \, \rho^{\frac{1}{2}}\, W_2(t_2-i\delta_2,x_{2}){\cal O}_j(b,x_j)  W_2(t_2,x_{2})  \, \rho^{\frac{1}{2}} \right\}}{ \text{Tr}\big\{ W_1W_1\,\rho \big\}\, \text{Tr}\big\{ {\cal O}_j(-a,x_j) \rho^{\frac{1}{2}}{\cal O}_j(b,x_j)\rho^{\frac{1}{2}}\big\} \,\text{Tr}\big\{ W_2 W_2\,\rho \big\}}
\end{equation}
where $W_{1,2}$ and ${\cal O}_j$ are operators with conformal dimensions $(h_{1,2},\bar h_{1,2})$ and $(h_j,\bar h_j)$. The parameters $\delta_{1,2}$ are infinitesimal and serve to regulate the result (we will sometimes leave these implicit for ease of notation). To compute this quantity, we will first discuss the eikonal action approach to conformal correlators, as discussed in \cite{Choi:2023mab} (which builds on previous work \cite{Shenker:2014cwa,Maldacena:2016upp} and especially \cite{Cotler:2018zff,Haehl:2018izb,Nguyen:2021jja}); afterwards, we will use boundary repametrizations to confirm the result. We will summarize the calculation in this section and leave the details to appendix \ref{appendix:details}. We give a third perspective based on analytic continuation of Euclidean conformal blocks in appendix \ref{sec:CFTblocks} We set $\beta=2\pi$ in this section.

\subsection{CFT six-point function from eikonal action}
\label{eikonal}

Following \cite{Choi:2023mab}, we discuss the eikonal action for reparametrizations in two-dimensional CFTs. Let us briefly recall how this method works. The goal is to compute the contribution of identity operator exchanges (and their descendants) to the six-point function \eqref{eq:F6def2} in a kinematic regime where the time differences between unequal operators are large, of order the scrambling time $\log c$. Generally, such identity exchanges can be generated by coordinate reparametrizations $(z,\bar{z}) \mapsto (\phi(z,\bar{z}),\bar{\phi}(z,\bar{z}))$,\footnote{ We suppress details here. See below for a more careful notation, taking into account the real-time contour.} since these source stress tensor insertions. The effective action for such reparametrizations is well known \cite{Alekseev:1988ce,Witten:1988} and can in fact be thought of as a generating functional for stress tensor correlation functions. The contribution of interest to \eqref{eq:F6def2} can thus be written as a real-time path integral over the reparametrizations $(\phi,\bar\phi)$. This effective action is proportional to $c$ and can therefore naively be approximated by a saddle point associated with the thermal state and small fluctuations around it. However, the OTOC \eqref{eq:F6def2} has the special feature that this saddle point approximation breaks down at late times: certain $SL(2,\mathbb{R})$ near-zero modes of the action (`scramblons') interact and yield an action that is exponentially small for large time separations. The large $c$ saddle point approximation thus breaks down and the {\it full} path integral over just these modes needs to be done. It is this `eikonal' contribution to the four-point function that dominates the behavior of our six-point function. We will now perform this path integral computation (delegating some details to appendix \ref{appendix:details}).

\begin{figure}
 \begin{center}                      
      \includegraphics[width=.7\textwidth]{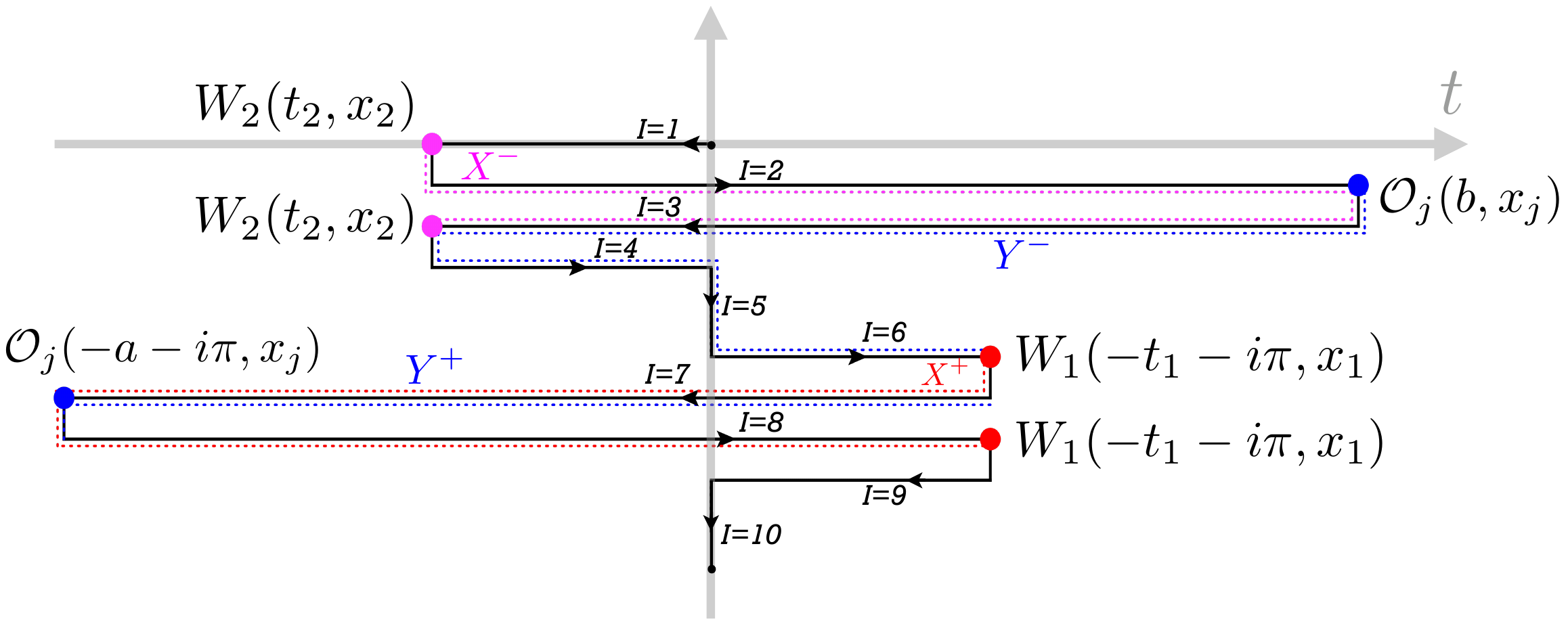}
      \caption{Generalized Keldysh contour appropriate for computing the six-point function. Euclidean contour separation is infinitesimal, except for $I=5$ and $I=10$, both of which have extent $-i\pi$. Endpoints are identified.}
  \label{fig:6contour}
  \end{center}
\end{figure}

We define the complex time contour for the path integral in terms of its parametrization as follows:
\begin{equation}
 t_I(s) = \left\{ \begin{aligned} 
  & -s & &(0\leq s \leq -t_{2}) \\
  & s+2t_{2} && ( -t_{2} \leq s \leq -2 t_{2}+b) \\
  &(-2t_{2}+2b)-s && (-2t_{2}+b \leq s \leq -3 t_{2}+2b ) \\
  & \qquad\vdots &&
  \end{aligned}
  \right.
  \label{eq:contourTime}
\end{equation}%
see \eqref{eq:contourTimeFull} for the complete parametrization. 
This parametrizes our six-fold contour as shown in figure \ref{fig:6contour}. We refer to the real parameter $s$ as `contour time'. The subscript $I=1,\ldots ,10$ labels the contour segment. We also use the following quantity to account for the correct signs on the various parts of the contour:
\begin{equation}
\label{eq:rsDef}
r_s\equiv i {dt_I (s)\over ds}=\begin{cases}i \qquad\quad \text{(}I=2,4,6,8\text{: `forwards')} \\ -i \qquad\; \text{(}I=1,3,7,9\text{: `backwards')}\\1  \qquad\quad \text{(}I=5,10\text{: `Euclidean')}\end{cases}
\end{equation}

The Alekseev-Shatashvili action for reparametrization modes $\phi_I(s,x)$ evaluated on this contour, reads as follows (see, e.g., \cite{Cotler:2018zff}):
\begin{equation}
\label{eq:SCFTnonlinear}
 iS = -\frac{c}{24\pi} \int ds dx\, \left\{r_s^3  \left[  \frac{\frac{1}{2}(\partial_s + ir_s \partial_x)  \partial_s \phi_I \;\, \partial_s^2 \phi_I}{\left( \partial_s \phi_I  \right)^2} + \frac{1}{2} (\partial_s + ir_s \partial_x)   \phi_I \;\, \partial_s \phi_I \right]  \, + \, \text{anti-holo.} \right\}
\end{equation}
where we omitted the second chiral copy for the anti-holomorphic reparametrization $\bar{\phi}_I(s,x)$.
The saddle point solution is $\phi_I(s,x) = t_I(s)+x$ and we will be interested in quadratic fluctuations $\delta \epsilon_I$ around this saddle.
The quadratic action for fluctuations is
\begin{equation}
\label{eq:SCFT}
 iS_{quad} = -\frac{c}{24\pi} \int ds dx\!  \left[ \frac{1}{2} (\partial_s + ir_s \partial_x) \delta \epsilon_I \;(r_s \partial_s^3 + r_s^3 \partial_s) \delta \epsilon_I 
 +
  \frac{1}{2} (\partial_s -ir_s \partial_x)  \delta \bar{\epsilon}_I \; (r_s\partial_s^3 + r_s^3 \partial_s) \delta \bar{\epsilon}_I\right] 
\end{equation}
where the contour time $s$ runs over its entire range \eqref{eq:contourTime}. The action has two copies of $x$-dependent $SL(2,\mathbb{R})$ symmetries. Relatedly, the following exponentially time-dependent zero modes are present: 
\begin{equation}
\label{eq:epsOTOCcft}
\begin{split}
SL(2,\mathbb{R}): \qquad \delta_+ \epsilon_I &= X^+(x) \,e^{-t_I(s)} \,,\qquad  \delta_- \epsilon_I = X^-(x) \,e^{t_I(s)} \,, \qquad \delta_0 \epsilon_I = X^0(x)
\end{split}
\end{equation}
and similarly for the anti-holomorphic part. We are particularly interested in the transformations $\delta_\pm$. They are infinitesimal versions of the following finite symmetry transformations:
\begin{equation}
SL(2,\mathbb{R}): \qquad \left\{ \begin{aligned}\; F_I &\;\longrightarrow\; \tilde{F}_I^+ \equiv F_I - \frac{(1-F_I)^2\,  \, 
X^+}{2-(1 - F_I) \,\, 
X^+}\\
F_I &\;\longrightarrow\; \tilde{F}_I^- \equiv F_I + \frac{(1+F_I)^2\, 
X^-}{2-(1 + F_I) \, \, 
X^-} \end{aligned}\right.
\end{equation}
where $F_I \equiv \tanh \frac{\phi_I}{2}$ and the transformed variables are related by $\tilde{F}_I^\pm = \tanh \frac{\phi_I + \delta_\pm \epsilon_I}{2}$.

The zero modes with exponential time dependence give the contribution to the correlator, which dominates for large times. Specifically, the following modes are excited by the operator insertions as shown in figure \ref{fig:6contour}:
\begin{equation}
\label{eq:epsOTOCcft}
\begin{split}
W_2:\quad& \quad\;\;\, \delta \epsilon^{(W_2)}_{I=2,3} = X^-(x) \,e^{t_I(s)
}\,,\\
W_1:\quad & \quad\;\;\, \delta \epsilon^{(W_1)}_{I=7,8} = -X^+(x) \, e^{-t_I(s)
}\,,\\
{\cal O}_j: \quad& \left\{ \begin{aligned}\delta \epsilon^{({\cal O}_j)}_{I=3} &= Y^+(x) \, e^{-t_I(s)
} \\
 \delta \epsilon^{({\cal O}_j)}_{I=4,5,6} &= Y^+(x) \, e^{-t_I(s)
 } -  Y^-(x) \, e^{t_I(s)
 }  \\
 \delta \epsilon^{({\cal O}_j)}_{I=7} &= -  Y^-(x) \, e^{t_I(s)
 } \end{aligned}\right.
\end{split}
\end{equation}
and similarly for the second copy.
Strictly speaking, $Y^+$ ($Y^-$) should also live on segment $I=7$ ($I=3$). However, we absorb this effect into an exponentially small redefinition of $X^+$ ($X^-$) since the coefficient of $Y^+$ ($Y^-$) at the junction between $I=6,7$ ($I=3,4$) is much smaller than the coefficient of $X^+$ ($X^-$) at the same point in time.
The full configuration is given by the following superposition:
\begin{equation}
\begin{split}
 \delta \epsilon^\text{otoc}_I &=  \chi_{I3}\,\big( Y^+ \, e^{
 -t_I}  + X^- \, e^{t_I
 } \big) - \chi_{I7}\, \big( Y^- \, e^{t_I
 } + X^+ \, e^{-t_I
 } \big)  \\
    &\quad +\chi_{I2} \, \big(X^- \, e^{t_I
    }\big) - \chi_{I8} \,\big( X^+ \, e^{-t_I
    } \big)+ (\chi_{I4} + \chi_{I5} + \chi_{I6} ) \big( Y^+ \, e^{-t_I
    } -  Y^- \, e^{t_I
    } \big) 
\end{split}
\end{equation}
where $\chi_{IJ}$ is the characteristic function on the different contour segments.
To derive the eikonal action for the six-point function setup, we evaluate $S_{quad}$ on the configuration $\delta \epsilon_I^\text{otoc}$. Note that this gives zero in the `bulk' of any of the contour segments. However, we get finite boundary terms at the contour turning points when derivatives act on the characteristic functions $\chi_{IJ}$.\footnote{One might worry about an ambiguity in these boundary terms, as these are precisely the locations where $r_s$ is discontinuous. However, explicit calculation shows that these particular boundary terms are always proportional to the fourth power, which satisfies $r_s^4 = 1$ everywhere.} This gives non-trivial boundary terms (in time) at $s=-2t_{2}+b$, $s=-3t_{2}+2b$, $s=-4t_{2}+2b+\pi-t_{1}$, and $s=-4t_{2}+2b+\pi - 2t_{1}+a$. We find:
\begin{equation}
\begin{aligned}
  iS_{eik} =\ & -\frac{ic}{6} \int dp \;  \left[ 
  (1+ip)\,Y^+(p)X^-(-p)   + 
  (1-ip)\,Y^-(p) X^+(-p) \right]+\text{anti.holo.}
  \end{aligned}
\end{equation}
where $X^{\pm}(x) = \int dp \, e^{ i p x}\,X^{\pm}(p)$ and similarly for $Y^\pm$.
The zeros at $p=\pm i$ are the telltale signature of this action being due to the `gravitational' exchange of stress tensors (and their descendants), see, e.g., \cite{Choi:2023mab}.

The coupling of the operators to the scramblon modes proceeds by reparametrization of three conformal two-point functions. Let us denote the conformal two-point functions as $\,{\cal G}^{h,\bar{h}}_{IJ}(t,x;t',x') \equiv \langle T_{\cal C} \, {\cal O}^{h,\bar h}_I(t,x) {\cal O}^{h,\bar h}_J(t',x') \rangle_\beta$, where $I,J$ label the contour segment on which the operators are inserted, and $T_{\cal C}$ denotes contour ordering. For the reparametrized (and normalized) two-point functions, we have
{\small
\begin{equation}
\begin{split}
  \frac{\delta_{Y_-}{\cal G}^{h_1,\bar{h}_1}_{68}}{{\cal G}^{h_1,\bar{h}_1}_{68}} &= \qty(\frac{1}{1+\frac{e^{-t_1}}{4i\sin\delta_1}\, Y^-(x_1)})^{2h_1}\qty(\frac{1}{1+\frac{e^{-t_1}}{4i\sin\delta_1}\, \bar{Y}^-(x_1)})^{2\bar{h}_1} \\
  \frac{\delta_{Y^+}{\cal G}^{h_2,\bar{h}_2}_{13}}{{\cal G}^{h_2,\bar{h}_2}_{13}} &= \qty(\frac{1}{1-\frac{e^{-t_2}}{4i\sin\delta_2}\,Y^+(x_2)})^{2h_2}\qty(\frac{1}{1-\frac{e^{-t_2}}{4i\sin\delta_2}\,\bar{Y}^+(x_2)})^{2\bar{h}_2} \\
  \frac{\delta_{X^\pm}{\cal G}^{h_j,\bar{h}_j}_{27}}{{\cal G}^{h_j,\bar{h}_j}_{27}} &=\qty(\frac{1}{1-\frac{e^{b}}{2(1+e^{a+b})} \, X^-(x_j) + \frac{e^{a}}{2(1+e^{a+b})} \, X^+(x_j)-\frac{e^{a+b}}{1+e^{a+b}}\frac{X^+(x_j)X^-(x_j)}{4}})^{2h_j}\\
  &\ \ \ \ \ \ \ \times\qty(\frac{1}{1-\frac{e^{b}}{2(1+e^{a+b})} \, \bar X^-(x_j) + \frac{e^{a}}{2(1+e^{a+b})} \, \bar X^+(x_j)-\frac{e^{a+b}}{1+e^{a+b}}\frac{\bar X^+(x_j)\bar X^-(x_j)}{4}})^{2\bar{h}_j}
\end{split}
\end{equation}}
The eikonal evaluation of the six-point function \eqref{eq:F6def2} then amounts to the following path integral over the `scramblon' modes:
\begin{equation}
\label{eq:F6calc1}
\begin{split} 
{\cal F}_{6,eik}\overline{\cal F}_{6,eik} &=\int \frac{dX^\pm dY^\pm }{{\cal Z}}\int \frac{d\bar X^\pm d\bar Y^\pm}{{\cal\bar Z}}\,  \frac{\delta_{Y^-}{\cal G}^{h_1,\bar{h}_1}_{68}}{{\cal G}^{h_1,\bar{h}_1}_{68}}
	\,\frac{\delta_{Y^+}{\cal G}^{h_2,\bar{h}_2}_{13}}{{\cal G}^{h_2,\bar{h}_2}_{13}} \,  \frac{\delta_{X^\pm}{\cal G}^{h_j,\bar{h}_j}_{27}}{{\cal G}^{h_j,\bar{h}_j}_{27}}  \;\\
	&\ \ \ \times\exp\Bigg\{-\frac{ic}{6} \int dp \;  \left[ 
  (1+ip)\,Y^+(p)X^-(-p)   + 
  (1-ip)\,Y^-(p) X^+(-p) \right]+\text{anti.holo.} \Bigg\}
\end{split}
\end{equation}
The normalization ${\cal Z}$ is chosen such that ${\cal F}_{6,eik}$ is normalized to 1 for coincident times. Similarly for ${\cal \bar Z}$. This integral can be performed explicitly and takes a simple form in the saddle point approximation where $h_{1,2} \gg h_j$ and $\bar h_{1,2} \gg \bar{h}_j$. We provide some details in appendix \ref{appendix:details} and quote the result here:
\begin{equation}
\begin{split}
 {\cal F}_{6,eik} &\approx  \bigg(1+\frac{3h_2}{2c\sin\delta_2}\, \frac{e^{b-t_2-|x_j-x_2|}}{1+e^{a+b}}\,\Theta(x_2-x_j)+\frac{3h_1}{2c\sin\delta_1}\, \frac{e^{a-t_1-|x_j-x_1|}}{1+e^{a+b}}\,\Theta(x_j-x_1)\\
	&\qquad\;\;  +\frac{9h_1h_2}{4c^2\sin\delta_1\sin\delta_2}\, \frac{e^{a+b}}{1+e^{a+b}}e^{-t_1-t_2-|x_1-x_2|}\,\Theta(x_j-x_1)\Theta(x_2-x_j)\bigg)^{-2h_j} \qquad\;\; (h_{1,2} \gg h_j) \\
\overline{{\cal F}}_{6,eik} &\approx \bigg(1+\frac{3\bar h_2}{2c\sin\delta_2}\, \frac{e^{b-t_2-|x_j-x_2|}}{1+e^{a+b}}\,\Theta(x_j-x_2)+\frac{3\bar h_1}{2c\sin\delta_1}\, \frac{e^{a-t_1-|x_j-x_1|}}{1+e^{a+b}}\,\Theta(x_1-x_j)\\
	&\qquad\;\;  +\frac{9\bar h_1\bar h_2}{4c^2\sin\delta_1\sin\delta_2}\,\frac{e^{a+b}}{1+e^{a+b}} e^{-t_1-t_2-|x_1-x_2|}\,\Theta(x_1-x_j)\Theta(x_j-x_2)\bigg)^{-2\bar h_j} \qquad\;\; (\bar h_{1,2} \gg \bar h_j)
\end{split}
\label{eq:F6eikResult}
\end{equation}
where we also wrote the analogous anti-holomorphic part.\footnote{ Note the step functions in space. Had we evaluated the 6-point function by analytic continuation of a Euclidean conformal block, then these discontinuities would have arisen from an exchange of dominance of different channels. See \cite{Mezei:2019dfv} for more discussion of this phenomenon, and appendix \ref{sec:CFTblocks} for explicit verification.}
 In these expressions we wish to take the limit where $a,b$ are much larger than any other time scale involved. Which terms can be dropped in this limit depends on the configuration of the spatial insertions:
 {\small
\begin{equation}
\label{eq:answer}
	\mathcal{F}_{6,eik}\,\overline{\mathcal{F}}_{6,eik} \approx \left\{\begin{aligned}
	& \qty(\frac{1}{1+\frac{9h_1h_2}{4c^2\sin\delta_1\sin\delta_2}e^{-t_1-t_2-|x_1-x_2|}})^{2h_j}
	\qquad && (x_1<x_j<x_2)  \\
	& \qty(\frac{1}{1+\frac{9{\bar h}_1{\bar h}_2}{4c^2\sin\delta_1\sin\delta_2}e^{-t_1-t_2-|x_1-x_2|}})^{2\bar{h}_j} 
	\qquad && (x_2<x_j<x_1)\\
	& \qty(\frac{1}{1+\frac{3h_1}{2c\sin\delta_1}\, e^{-t_1-b-|x_j-x_1|}})^{2h_j} \qty( \frac{1}{1+\frac{3\bar h_2}{2c\sin\delta_2}\, e^{-a-t_2-|x_j-x_2|}})^{2\bar h_j} \approx 1
	\quad &&\quad\;\; (x_{1,2}<x_j)\\
	& \qty(\frac{1}{1+\frac{3\bar h_1}{2c\sin\delta_1}\, e^{-t_1-b-|x_j-x_1|}})^{2\bar h_j} \qty( \frac{1}{1+\frac{3 h_2}{2c\sin\delta_2}\, e^{-a-t_2-|x_j-x_2|}})^{2 h_j} \approx 1
	\quad &&\quad\;\; (x_j<x_{1,2})
	 \end{aligned}\right.
\end{equation}
}
for $h_{1,2} \gg h_j$ and $a,b \gg -t_{1,2}$.
\begin{figure}
 \begin{center}                      
      \includegraphics[width=3.5
      in]{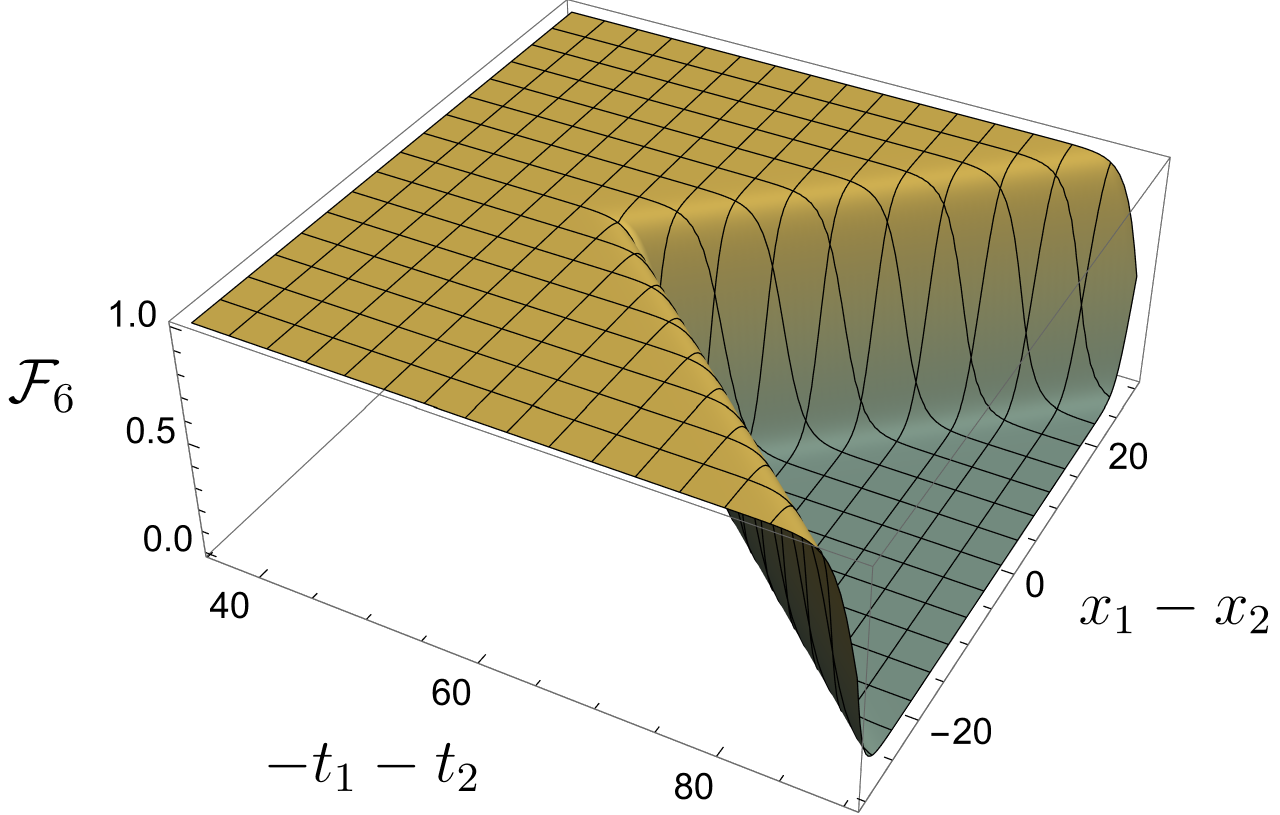}
      \caption{Plot of $\mathcal{F}_6(x)$ when $x$ is between $x_1$ and $x_2$. We chose $t_{*1}+t_{*2} = 60$, $h_j = \bar h_j = \frac{1}{2}$.}
  \label{plot}
  \end{center}
\end{figure}
Let $x_j = x$ in \eqref{eq:F6def2}, then \eqref{eq:F6def2} reduces to $\mathcal{F}_6(x)$ we proposed to diagnose the black hole formation in section \ref{sec:diagnosis}. Figure \ref{plot} is a plot of $\mathcal{F}_6(x)$ when $x$ is between $x_1$ and $x_2$, as a function of $-t_1-t_2$ and $x_1-x_2$. As explained in section \ref{sec:diagnosis}, we see a sharp transition when $\mathcal{F}_6$ drops from $1$ to $0$. The transition happens when $-t_1-t_2-|x_1-x_2|\approx t_{*1}+t_{*2}$, i.e., when the butterfly cones start to overlap.

\subsection{Computation of six-point function via boundary reparametrizations}
\label{sec:boundaryReparam}

In the limit when $h_{1,2}\geq h_j$ and $\bar h_{1,2}\geq \bar h_j$, the six-point function \eqref{eq:F6def2} can be considered as a correlator between two operators ${\cal O}_j^L(a,x_j)$ and ${\cal O}_j^R(b,x_j)$ on a background with two perturbations created by $W_1^L(t_1,x_{1})$ and $W_2^R(t_{2},x_{2})$. We can thus compute the correlator by solving for the backreaction due to the operator insertions $W_1^L(t_1,x_{1})$ and $W_2^R(t_{2},x_{2})$.

To warm up, we first recall the analogous calculation in the Schwarzian theory describing $AdS_2$ gravity. The backreaction is equivalent to repametrizations of the left and right boundary times $t_{R,L} \longrightarrow \tilde{t}_{R,L}$, given by
\begin{align}
\label{repametrization_1}
    &e^{\tilde t_L} = \frac{e^{t_L}}{1+\alpha_1e^{t_L}}\,,\ \ \ \ e^{\tilde t_R} = \frac{e^{t_R}}{1+\alpha_2 e^{t_R}} \,,
\end{align}
where $\alpha_1 = \frac{\Delta_1}{C}e^{-t_{1}}$ and $\alpha_2 = \frac{\Delta_2}{C}e^{-t_{2}}$. It is easy to check that this boundary repametrization is equivalent to the bulk statement that the horizons have been shifted outwards by amounts $\alpha_{1,2}$ in Kruskal coordinates. With these reparametrizations, the probe operator transforms as
\begin{align}
	\psi_j\longrightarrow\tilde\psi_j(t) = \tilde t'(t)^{\Delta_j}\psi_j(\tilde t(t))
\end{align}
The correlator transforms as
\begin{equation}
\begin{aligned}
	\frac{\expval{\tilde\psi_j^L(t_L) \tilde\psi_j^R(t_R)}}{\expval{\psi_j^L(t_L)\psi_j^R(t_R)}} &= \frac{\qty(\tilde t_R'(t_R))^{\Delta_j}\qty(\tilde t_L'(t_L))^{\Delta_j}}{\qty[\cosh(\frac{\tilde t_R+\tilde t_L}{2})]^{2\Delta_j}}\qty[\cosh(\frac{t_R+t_L}{2})]^{2\Delta_j}\\
	 &= \qty(\frac{1}{1+\frac{e^{t_L}}{1+e^{t_L+t_R}}\,\alpha_1+\frac{e^{t_R}}{1+e^{t_L+t_R}}\,\alpha_2+\frac{e^{t_L+t_R}}{1+e^{t_L+t_R}}\,\alpha_1\alpha_2})^{2\Delta_j} \,.
\end{aligned}
\end{equation}
With $t_L = a$, $t_R = b$ and taking $a$, $b$ large, we have
\begin{align}
	\frac{\expval{\psi_1^L(t_{1})\psi_2^R(t_{2})\psi_j^L(a) \psi_j^R(b)\psi_1^L(t_{1})\psi_2^R(t_{2})}}{\expval{\psi_j^L(a)\psi_j^R(b)}} = \qty(\frac{1}{1+\frac{\Delta_1\Delta_2}{C^2}e^{-t_{1}-t_{2}}})^{2\Delta_j}\,.
\end{align}
This also agrees with the geodesic calculation in $AdS_2$. 

Coming back to two-dimensional CFTs, we have a holomorphic part and an anti-holomorphic part, each of which works in a way similar to the Schwarzian. We let $x^+ \equiv t+x$, $x^- \equiv t-x$.
Focusing on the right perturbation at $(t_{2},x_2)$, we find that it induces the following boundary reparametrization $(t_R,x_R) \longrightarrow (\tilde{t}_R, \tilde{x}_R)$:
\begin{equation}
\label{repametrization_2}
\begin{aligned}
	&e^{\tilde x_R^+} = e^{\tilde t_R+\tilde x_R} = \frac{1}{1+\theta(-(x_R-x_{2}))\frac{3h_2}{2c\sin\delta_2}e^{x^+_R-x^+_{2}}}\\
	&e^{\tilde x^-_R} = e^{\tilde t_R-\tilde x_R} = \frac{1}{1+\theta(x_R-x_{2})\frac{3\bar h_2}{2c\sin\delta_2}e^{x^-_R-x^-_{2}}}
\end{aligned}
\end{equation}
Note that this particular reparametrization is also obtained as the saddle point value of $X^-(x)$ defined in section \ref{eikonal} (see \eqref{eq:Xmresult}).
Note further the similarity between \eqref{repametrization_1} and \eqref{repametrization_2}. 

From now on we assume all operators are scalars, i.e., $h = \bar h = \frac{\Delta}{2}$. 
In appendix \ref{app:boundary_rep} we show that the boundary repametrization \eqref{repametrization_2} is equivalent to a bulk horizon shift given by $h(x)$ as in \cite{Roberts:2014isa}. Carrying out the computation of the six-point function similar to Schwarzian case, taking $a,b$ large, and assuming $x_j$ is in between $x_{1}$ and $x_{2}$, we again get
\begin{equation}
\label{result_2}
\begin{split}
	&\frac{\expval{W_1^L(t_1,x_{1})W_2^R(t_2,x_{2}){\cal O}_j^L(a,x_j) {\cal O}_j^R(b,x_j)W_1^L(t_{1},x_{1})W_2^R(t_{2},x_{2})}}{\expval{W_1W_1}\expval{{\cal O}_j^L(a,x_j){\cal O}_j^R(b,x_j)}\expval{W_2W_2}} 
	\\
	=\ &\qty(\frac{1}{1+\frac{9\Delta_1\Delta_2}{16c^2\sin\delta_1\sin\delta_2}e^{-t_{1}-t_{2}-|x_{1}-x_{2}|}})^{\Delta_j}
\end{split}
\end{equation}

One can also carry out similar computations when $x_j>x_{1,2}$ or $x_j<x_{1,2}$. We get same result as in \eqref{eq:answer} for a scalar perturbation.

One curious fact is that \eqref{result_2} no longer agrees with the (naive) bulk geodesic calculation. The reason is that the boundary repametrization \eqref{repametrization_2} is not continuous at $x_R=x_{2}$, which removes a wedge from $AdS_3$. Indeed, from the transformation of correlators under reparametrizations, one finds that the calculation \eqref{result_2} does give the same result as a geodesic calculation in $AdS_3$, but the correct bulk geometry is not exactly $AdS_3$ but rather $AdS_3$ with appropriate wedges removed.

\section{Discussion}
\label{sec:discussion}

In this paper we studied the process of two high energy particles colliding in $AdS$. Treating AdS as a pair of entangled hyperbolic black holes, we describe the collision process in terms of a quantum circuit. Using this picture, we proposed a threshold condition for the formation of a black hole: when the butterfly cones of the two perturbations overlap in the shared quantum circuit, the collision will create a black hole in $AdS$. This proposal connects operator growth with black hole formation. 

There are various unanswered questions. We confirmed the correctness of the threshold condition in three bulk dimensions. One would also like to understand the black hole formation transition in a dual two-dimensional conformal field theory. In particular, one can consider the four-point function built out of pairs of operators setting up the collision process, and analyze the decomposition into conformal blocks in different channels. To identify the proposed black hole formation threshold, it would be interesting to look for an exchange of dominance of different channels, depending on the kinematic setup.

Before we discussed CFTs, we argued for the SYK chain six-point function \eqref{6_point_function} as a good diagnostic for the black hole formation transition. It would be very worthwhile to numerically compute this quantity by simulating the SYK chain and confirming the expected features. Such numerical simulations have been done for simpler cases, such as the four-point OTOC in complex SYK \cite{Fu:2016yrv}, in Brownian SYK \cite{Sunderhauf:2019djv}, and in standard SYK \cite{Kobrin:2020xms}. State of the art methods in \cite{Kobrin:2020xms} achieve results for $N \sim 60$ fermions. We suspect that this is still insufficient for obtaining good results in the SYK chain (with several sites), not to mention six-point OTOCs. We therefore leave this challenge for future analysis.

The situation in higher dimensions is more interesting as well as more subtle. We expect our threshold condition to hold when the impact parameter $b$ is larger than $\ell_{AdS}$. When $b$ is less than $\ell_{AdS}$, a small black hole may form, which likely cannot be captured by the quantum circuit picture as it may only resolve processes on super-$AdS$ scales. On the other hand, it is interesting to contemplate if the CFT six-point function we used can diagnose the formation of small black holes in higher dimensions. To answer this question one needs a more detailed understanding of the behavior of various out-of-time-ordered correlation functions at small distances in higher dimensions. We leave this to future work.  

It is also desirable to obtain a deeper physical understanding of the connection between operator growth and black hole formation. The logic behind our proposal is simple: as we observe a sharp transition in the quantum circuit, we expect a dramatic transition in the bulk -- a natural guess is black hole formation,\footnote{There may also be naked singularity formations in higher dimensions. We thank Netta Engelhardt for pointing this out.} and this is indeed correct at least in low dimensions. However, we do not have a deep understanding of why such a connection exists. Perhaps more interestingly, can we gain a microscopic understanding of the black hole formation process? For this purpose, one could study more concrete theories like D1-D5 system at the orbifold point \cite{Balasubramanian:2016ids}.\footnote{We thank Vijay Balasurbramnian for pointing this out.} And, finally, can we elucidate the formation of the black hole singularity? We leave these questions for future work.

\section*{Acknowledgements}
We thank Vijay Balasubramanian, Netta Engelhardt, Daniel Harlow, Hong Liu, Juan Maldacena, Mark Mezei for helpful discussions and comments. F.H.\ is supported by the UKRI Frontier Research Guarantee [EP/X030334/1].
Y.Z.\ was supported in part by the National Science Foundation under Grant No. NSF PHY-1748958 and by a grant from the Simons Foundation (815727, LB).

\appendix

\section{Details on CFT calculations}
 \label{appendix:details}

This appendix provides some details for section \ref{sec:calculations}.

\subsection{Eikonal integral in 2d CFT}

Here we provide some calculational details complementing section \ref{eikonal}. The complete parametrization of the contour is
{\small
\begin{equation}
 t_I(s) = \left\{ \begin{aligned} 
  & -s & &(0\leq s \leq -t_{2}) \\
  & s+2t_{2} && ( -t_{2} \leq s \leq -2 t_{2}+b) \\
  &(-2t_{2}+2b)-s && (-2t_{2}+b \leq s \leq -3 t_{2}+2b ) \\
  &s-(-4t_{2}+2b) && (-3 t_{2}+2b \leq -4 t_{2}+2b) \\
  &-i(s-(-4t_{2}+2b)) && (-4t_{2}+2b \leq s \leq -4t_{2}+2b+ \pi)\\
  &-i\pi + s - (-4t_{2}+2b+\pi) && (-4t_{2}+2b+\pi \leq s \leq -4t_{2}+2b+ \pi -t_{1}) \\
  &-i\pi + (-4t_{2}+2b+\pi-2t_{1})-s && (-4t_{2}+2b+\pi -t_{1}\leq s \leq -4t_{2}+2b+ \pi - 2t_{1}+a) \\
  &-i\pi +s- (-4t_{2}+2b+\pi-2t_{1}+2a) && (-4t_{2}+2b+ \pi - 2t_{1}+a \leq s \leq -4t_{2}+2b+ \pi - 3t_{1}+2a) \\
  &-i\pi + (-4t_{2}+2b+\pi-4t_{1}+2a) - s && (-4t_{2}+2b+ \pi - 3t_{1}+2a \leq s \leq -4t_{2}+2b+ \pi -4t_{1}+2a) \\
  &-i( s-(-4t_{2}+2b-4t_{1}+2a)) && (-4t_{2}+2b+ \pi - 4t_{1}+2a \leq s \leq -4t_{2}+2b+ 2\pi - 4t_{1}+2a) 
  \end{aligned}
  \right.
  \label{eq:contourTimeFull}
\end{equation}
}

We shall now give some details on the evaluation of the scramblon path integral \eqref{eq:F6calc1}. 
The equal-time vertex functions appearing in it can be expressed in terms of null momentum variables:
{\small
\begin{equation}
\begin{split}
	\qty(\frac{1}{1+\frac{e^{-t_1}}{4i\sin\delta_1}\, Y^-(x_1)})^{2h_1}&= \frac{(2\sin\delta_1)^{2h_1}}{\Gamma(2h_1)}\int_{-\infty}^0 dp_-\exp(-i\int dp\, e^{ipx_1}Y^-(p)p_-)\frac{(-2p_-e^{t_1})^{2h_1}}{-p_-}e^{4p_-e^{t_1}\sin\delta_1}\\
	\qty(\frac{1}{1-\frac{e^{-t_2}}{4i\sin\delta_2}\,Y^+(x_2)})^{2h_2} &= \frac{(2\sin\delta_2)^{2h_2}}{\Gamma(2h_2)}\int_{-\infty}^0 dp_+\exp(i\int dp\, e^{ipx_2}Y^+(p)p_+)\frac{(-2p_+e^{t_2})^{2h_2}}{-p_+}e^{4p_+e^{t_2}\sin\delta_2}
\end{split}
\end{equation}
}%
Plugging these expressions into \eqref{eq:F6calc1}, the $Y^\pm(p)$ integrals lead to delta-function constraints, which fix $X^\pm(p)$ in terms of the lightcone momenta:
\begin{align}
	\frac{c}{6}(1-ip)X^+(-p) &= -p_-e^{ipx_1}\\
	\Rightarrow \; X^+(x) 
	&= -\frac{6}{ c}\,p_-\int \frac{dp}{2\pi}\, \frac{e^{ip(x-x_1)}}{1+ip} = -\frac{6}{c}\,p_-\,e^{-|x-x_1|}\Theta(x-x_1)\\
	\frac{c}{6}(1+ip)X^-(-p) &= e^{ipx_2}\,p_+\\
	\Rightarrow \; X^-(x) &= \frac{6}{c}\,p_+\int \frac{dp}{2\pi} \, \frac{e^{ip(x-x_2)}}{1-ip} = \frac{6}{c}\,p_+\,e^{-|x-x_2|}\Theta(x_2-x)
	\label{eq:Xmresult}
\end{align}
These constraints trivialize the $X^\pm$ integrals and we find:
{\small
\begin{equation}
\begin{split}
	\mathcal{F}_{6,eik} &\propto  \int_{-\infty}^0 dp_-\frac{(-4p_-\sin\delta_1\, e^{t_1})^{2h_1}}{-p_-}\, e^{4p_-e^{t_1}\sin\delta_1}\int_{-\infty}^0 dp_+\frac{(-4p_+\sin\delta_2\, e^{t_2})^{2h_2}}{-p_+}\,e^{4p_+e^{t_2}\sin\delta_2}\\
	&\qquad\qquad \times \bigg(1- \frac{3}{c} \, p_+ \, \frac{e^{b-|x_j-x_2|}}{1+e^{a+b}} \Theta(x_2-x_j)- \frac{3}{c} \, p_- \, \frac{e^{a-|x_j-x_1|}}{1+e^{a+b}} \Theta(x_j-x_1) \\
	&\qquad\qquad\qquad\;\; +\frac{9}{c^2}p_+p_-\frac{e^{a+b}}{1+e^{a+b}}e^{-|x_j-x_1|-|x_j-x_2|}\,\Theta(x_j-x_1)\Theta(x_2-x_j)\bigg)^{-2h_j}\\
	&= \int_0^{\infty}d\tilde p_-\int_0^{\infty}d\tilde p_+\tilde p_-^{2h_1-1}\tilde p_+^{2h_2-1}e^{-\tilde p_--\tilde p_+} 
	 \bigg(1+\frac{3}{c}\frac{\tilde p_+}{4\sin\delta_2}\, \frac{e^{b-t_2-|x_j-x_2|}}{1+e^{a+b}}\,\Theta(x_2-x_j)\\
	&\qquad+\frac{3}{c}\frac{\tilde p_-}{4\sin\delta_1}\, \frac{e^{a-t_1-|x_j-x_1|}}{1+e^{a+b}}\,\Theta(x_j-x_1)  +\frac{9}{c^2}\frac{\tilde p_+\tilde p_-}{16\sin\delta_1\sin\delta_2}\, \frac{e^{a+b}}{1+e^{a+b}}e^{-t_1-t_2-|x_1-x_2|}\,\Theta(x_j-x_1)\Theta(x_2-x_j)\bigg)^{-2h_j}
	\end{split}
\end{equation}}
where we simplified the spatial dependence using the step functions.
When $h_{1,2} \gg h_j$ the remaining integrals can be evaluated by a saddle point approximation, with saddle point values $\tilde p_- = 2h_1$, $\tilde p_+ = 2h_2$. This yields the expression \eqref{eq:F6eikResult} in the main text.

\subsection{$AdS_3$ Boundary reparametrizations}
\label{app:boundary_rep}

\begin{figure}
 \begin{center}                      
      \includegraphics[width=4
      in]{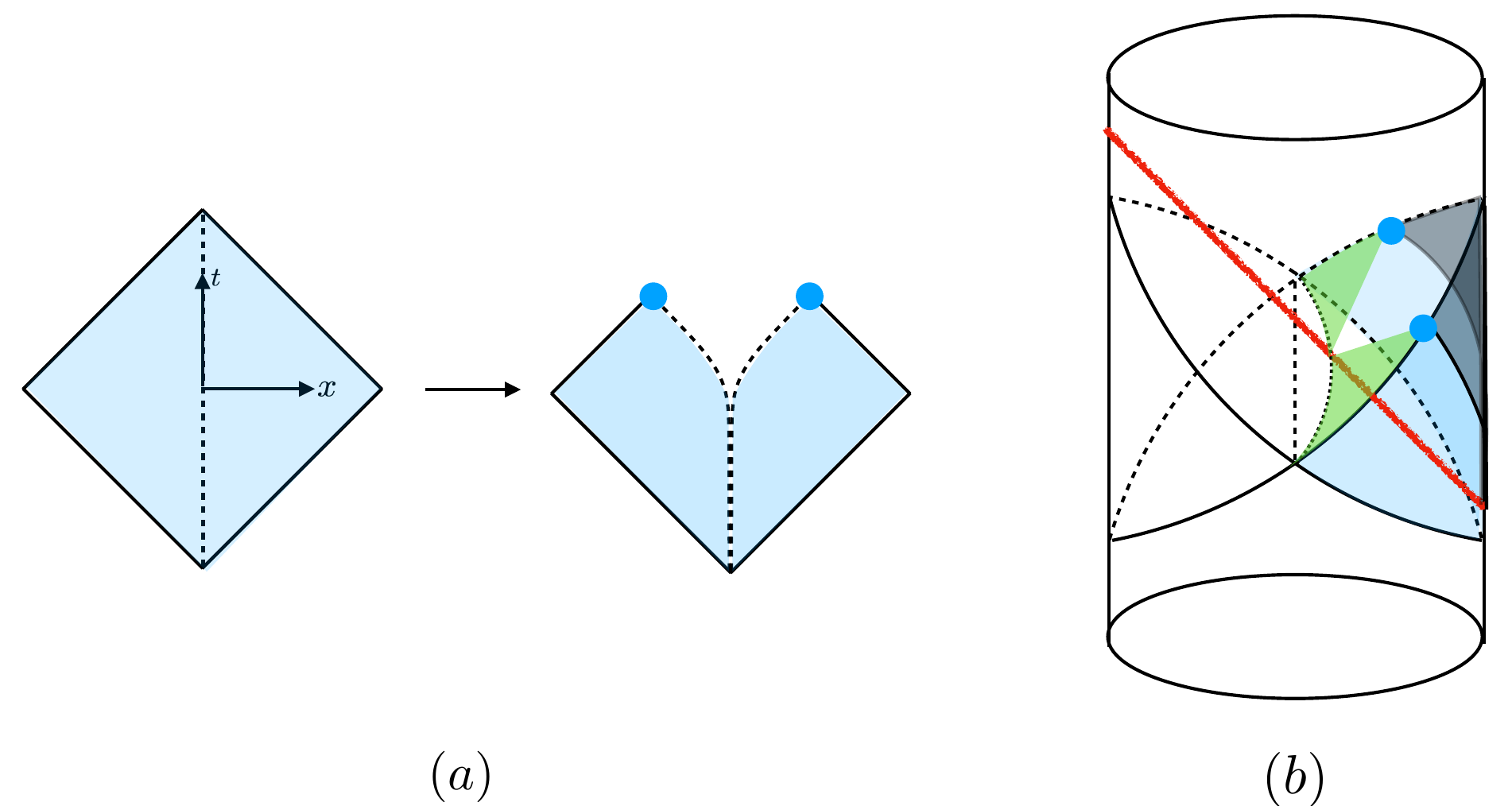}
      \caption{Boundary reparametrization describing the effect of the perturbation: a wedge is removed from the bulk geometry and the reparametrized boundary is discontinuous across $x=0$.}
  \label{fig_reparametrization}
  \end{center}
\end{figure}

In this appendix we develop a more detailed understanding of the boundary repametrizations in \eqref{repametrization_2}. Let us assume we have a scalar perturbation, so $h = \bar h$. We further set $x_2 = 0$. Notice that this reparametrization is discontinuous across $x = 0$. Equivalently, a wedge near $x=0$ is removed as shown in figure \ref{fig_reparametrization}(a).   
This discontinuity is physical: in the bulk, we need to remove a wedge from global $AdS_3$ after the particle enters, as the particle will create a conical defect and the bulk is no longer global $AdS_3$. 

Next, we will show that this construction is equivalent to a shift at the horizon by $h(x) = \frac{3h}{2c\sin\delta_2}e^{-t_2-|x|}$. 
Let $\alpha = \frac{3h}{2c\sin\delta_2}e^{-t_2}$. We write the repametrization as
\begin{equation}
\label{reparametrization_3}
\begin{aligned}
	&\tilde t+ |\tilde x| = t+|x| \,,
	&e^{\tilde t-|\tilde x|} = \frac{e^{t-|x|}}{1+\alpha e^{t-|x|}}\,.
\end{aligned}
\end{equation}
For $AdS_3$, the transformation from Poincar\'{e} to embedding coordinates is given by
\begin{align}
	Y^{-1} = \frac{t_P}{z}\,,\ \ Y^0 = -\frac{t_P^2-z^2-y_P^2-1}{2z}\,,\ \ Y^1 = \frac{t_P^2-z^2-y_P^2+1}{2z}\,,\ \ Y^2 = -\frac{y_P}{z}\,.
\end{align}
Similarly, the transformation from Rindler to embedding coordinates is given by
\begin{align}
	Y^{-1} = \sqrt{r^2-1}\sinh(t)\,,\ \ Y^0 = r\cosh(x)\,,\ \ Y^1 = \sqrt{r^2-1}\cosh(t)\,,\ \ Y^2 = -r\sinh(x)\,.
\end{align}

Now consider the blue point in figure \ref{fig_reparametrization}(b). It satisfies $\tilde t+|\tilde x| = \infty$, $\tilde t-|\tilde x| = -\log(\alpha)$, and we consider the intersection of its past light cone (green sheets in figure \ref{fig_reparametrization}(b)) with the past horizon. We focus on the one with $x>0$. In Poincar\'{e} coordinates, this locus is given by
\begin{align}
	y_P^+\equiv t_P+y_P = 1\,, \qquad y_P^-\equiv t_p-y_P = \tanh(-\frac{1}{2}\log\alpha) = \frac{1-\alpha}{1+\alpha}\,.
\end{align}
Consider null geodesics originating from this point. In Poincar\'{e} coordinates, these are defined by the following Lagrangian: 
\begin{align}
	L = \frac{\dot y_P^+\dot y_P^--\dot z^2}{z^2}
\end{align}
From this we deduce conserved quantities:
\begin{align}
	p_+ = \frac{\dot y_P^-}{z^2}\,,\ \ p_- = \frac{\dot y_P^+}{z^2}\,,
\end{align}
such that
\begin{align}
	\frac{dz}{d\tau} &= z^2 \sqrt{p_+p_-}\,,\\
	\frac{dy_P^-}{dz} &= -\sqrt{\frac{p_+}{p_-}}\,,\ \ \frac{dy_P^+}{dz} = -\sqrt{\frac{p_-}{p_+}} \,.
\end{align}
From this we find for the geodesic:
\begin{align}
\label{eq:yP}
	y_P^+ = 1-z\sqrt{\frac{p_-}{p_+}}\,, \ \ \ y_P^- = \frac{1-\alpha}{1+\alpha}-z\sqrt{\frac{p_+}{p_-}}\,.
\end{align}
The past horizon is given by $Y^{-1}+Y^1 = 0$, which is equivalent to $(t_P+1)^2-z^2-y_P^2 = 0$, i.e., $(y_P^++1)(y_P^-+1) = z^2$. Using \eqref{eq:yP}, the intersection of the geodesic with the past horizon is parametrized by:
\begin{align}
	\qty(1-z\sqrt{\frac{p_-}{p_+}}+1)\qty(\frac{1-\alpha}{1+\alpha}-z\sqrt{\frac{p_+}{p_-}}+1) = z^2 \,.
\end{align}
Let $\lambda\equiv \sqrt{\frac{p_+}{p_-}}$. We can then write:
\begin{align}
	z = \frac{2\lambda}{1+(1+\alpha)\lambda^2}\,,\ \ y_P^+ = 1-\frac{2}{(1+\alpha)\lambda^2+1}\,,\ \ y_P^- = -1+\frac{1}{1+\alpha}\frac{2}{1+(1+\alpha)\lambda^2}\,.
\end{align}
This is the location of the new event horizon in Poincar\'{e} coordinates. 
Going back to Rindler coordinates, we can write this as follows:
\begin{align}
 -\frac{Y^2}{Y^0} &= \tanh(x) = -\frac{2y}{t_P^2-z^2-y_P^2-1} = -\frac{y_P^+-y_P^-}{y_P^+y_P^--z^2-1} = \frac{(1+\alpha)^2\lambda^2-1}{(1+\alpha)^2\lambda^2+1}\\
	\Leftrightarrow \quad x &= \log((1+\alpha)\lambda) \quad \Leftrightarrow \quad e^{-x} = \frac{1}{(1+\alpha)\lambda} \,.
\end{align}
Note that we assumed $x>0$, which requires $\lambda>\frac{1}{1+\alpha}$. 
Similarly, the Kruskal coordinate of the intersection is given by 
\begin{align}
	u = Y^{-1}-Y^1 = \frac{2t_p}{z} = \frac{y_P^++y_P^-}{z} = -\frac{\alpha}{(1+\alpha)\lambda} = -\alpha e^{-x} = -h(x)
\end{align}
This is exactly the correct shift along the past horizon, as we set out to show.

\section{CFT six-point function from analytic continuation}
\label{sec:CFTblocks}

In this appendix we confirm the eikonal result \eqref{eq:answer} using a different method: we start from the Euclidean Virasoro six-point identity block and analytically continue it to get the appropriate kinematic configuration. The six-point block at ${\cal O}(c^{-2})$ was worked out in \cite{Anous:2020vtw}. There, it was concluded that it takes the following form (adapted to our setup): 
\begin{equation}
\begin{split}
 {\cal F}_6 &= \frac{\text{Tr}\left\{ W_1 {\cal O}_j W_1 \rho^{\frac{1}{2}} {\cal O}_j  \rho^{\frac{1}{2}}\right\}}{\text{Tr}\left\{ W_1 W_1 \rho \right\} \text{Tr} \left\{ {\cal O}_j  \rho^{\frac{1}{2}}{\cal O}_j  \rho^{\frac{1}{2}}\right\}}
 \frac{\text{Tr}\left\{  {\cal O}_j  \rho^{\frac{1}{2}}W_2 {\cal O}_j W_2  \rho^{\frac{1}{2}}\right\}}{\text{Tr}\left\{ W_2 W_2 \rho \right\} \text{Tr} \left\{ {\cal O}_j  \rho^{\frac{1}{2}}{\cal O}_j  \rho^{\frac{1}{2}}\right\}}
 \frac{\text{Tr}\left\{  W_1 W_1 \rho^{\frac{1}{2}}W_2 W_2  \rho^{\frac{1}{2}}\right\}}{\text{Tr}\left\{ W_1 W_1 \rho \right\} \text{Tr} \left\{ W_2W_2 \rho\right\}} \\
 &\qquad \times \left( 1 + {\cal G}_T^\text{(6,star)} + {\cal U}_T^\text{(6,ext.)} + \ldots \right)\left( 1 + \overline{{\cal G}}_{\bar T}^\text{(6,star)} + \overline{{\cal U}}_{\bar T}^\text{(6,ext.)} + \ldots \right) \,.
\end{split}
\end{equation}
The right hand side of the first line simply consists of (conventional) four-point functions. In the second line each chiral factor has two contributions at leading order: a global $SL(2,\mathbb{R})$ six-point block in the `star' channel ${\cal G}_T^\text{(6,star)}$, and an `extra' contribution ${\cal U}_T^\text{(6,ext.)}$, which contributes at the same order to the Virasoro block.\footnote{ The `star' channel is also often referred to as the `snowflake' channel. It was studied also in, e.g., \cite{Jepsen:2019svc,Fortin:2020yjz,Hoback:2020pgj}.} In \cite{Anous:2020vtw} it was concluded that the global block does not give a relevant contribution to the OTOC.\footnote{ The precise OTOC considered in that reference was slightly different. However, one can check that the same holds in our case at hand.} We will therefore not concern ourselves with its explicit form. The detailed expression for the `extra' contribution in the Virasoro identity block is as follows:
\begin{equation}
	{\cal U}_T^\text{(6,ext.)} \equiv  \frac{18 h_1 h_2 h_j}{c^2} \left[ \widetilde{\cal I}(z,u,v) + \widetilde{\cal I}(z,v,u) + \widetilde{\cal I} \left( \frac{1}{z},\frac{u}{z}, \frac{v}{z} \right)   + \widetilde{\cal I} \left( \frac{1}{z},\frac{v}{z}, \frac{u}{z} \right)\right] \,,
\end{equation}
where
{\small
\begin{equation}
\label{eq:Itilde}
\begin{split}
\widetilde{\cal I}(z,u,v)
&=  \left[ \frac{2(2+u+v)}{1-z}-\frac{1+2u^2}{1-u} - \frac{1}{1-v}  - \frac{(u-v)z}{(z-u)(z-v)} - \frac{2u(v+(2+u)z)}{z(u-v)} - \frac{8uv\, \log z}{(u-v)(1-z)}  \right]\log u \\
&\quad - \frac{4(1-u)}{(u-v)(1-z)} \left[ 1+\frac{ vz-u^2}{u}-vz+2(z-v)   \right. \\
&\qquad\qquad\qquad\qquad\quad\;\;\, \left. + \frac{4(1-v)z}{1-z} \log z + \frac{2(uv-z)}{(1-u)} \log u- \frac{4(z-u)v}{u-v} \, \log \frac{u}{v} \right] \log(1-u)\,.
\end{split}
\end{equation}
}
Here, the three independent conformal cross ratios are 
 \begin{equation}
\label{eq:crossDef}
   z = \frac{(z_1-z_j)(z_j'-z_1')}{(z_1-z_j')(z_j-z_1')} \,, \qquad u = \frac{(z_1-z_j)(z_j'-z_2)}{(z_1-z_j')(z_j-z_2)} \,, \qquad v =  \frac{(z_1-z_j)(z_j'-z_2')}{(z_1-z_j')(z_j-z_2')}\,,
\end{equation}
where the insertion points of operators are denoted as $(z_1,z_1')$ for $W_1$ operators, $(z_2,z_2')$ for $W_2$ operators, and $(z_j,z_j')$ for ${\cal O}_j$ operators (and similarly for the anti-holomorphic coordinates). 

\begin{figure}
 \begin{center}                      
      \includegraphics[width=2.5in]{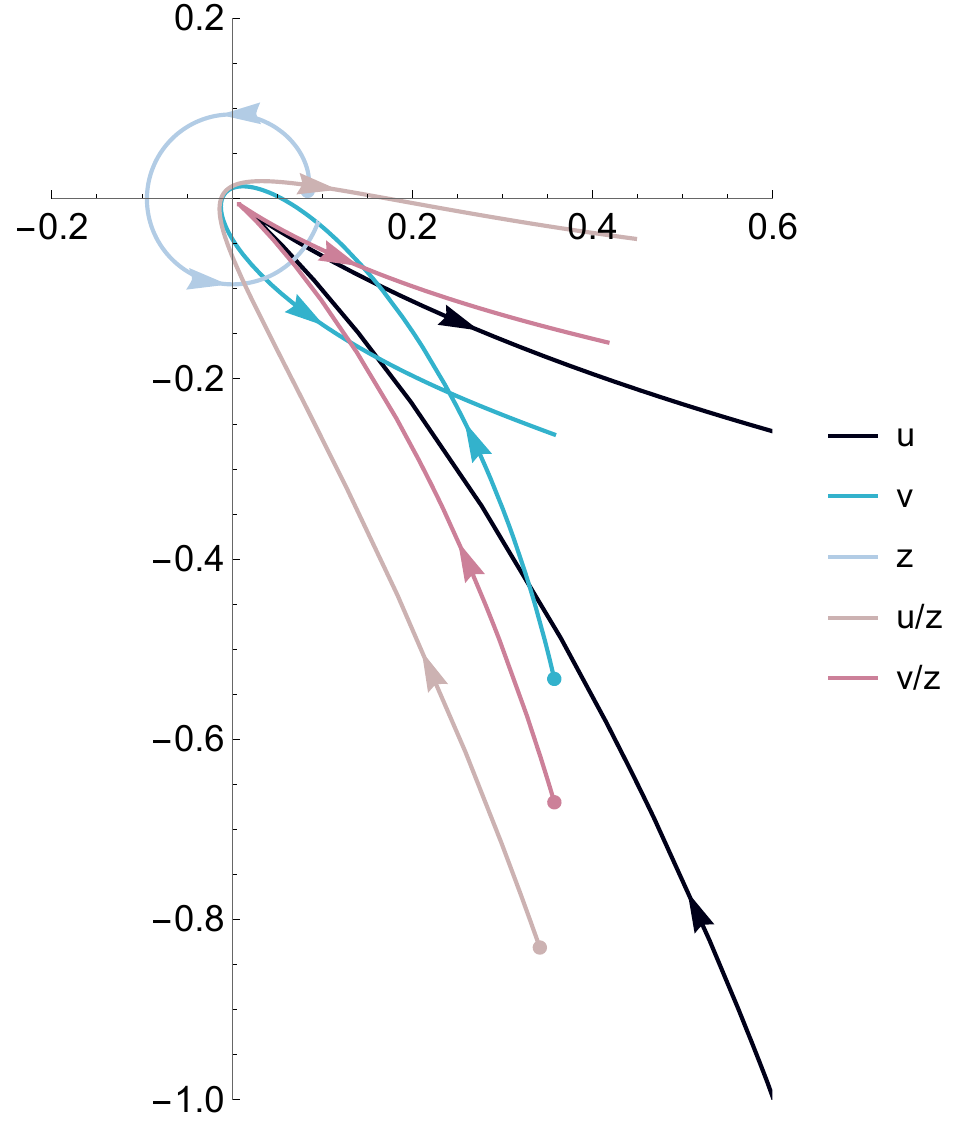}
      \caption{Analytic continuation of the cross ratios from the Euclidean section to the out-of-time-order configuration \eqref{eq:F6def2}. Here we take $(x_1,x_j,x_2) = (1,1.5,2)$ and $a=b=-10\, t_1 = -10\,t_2$.}
  \label{fig:paths}
  \end{center}
\end{figure}

The above expressions are in Euclidean signature. To obtain a contribution to our real-time correlator \eqref{eq:F6def2}, we analytically continue the Euclidean locations to suitable Lorentzian times. In this process the cross ratios may or may not cross the branch cuts of the logarithms in \eqref{eq:Itilde}. Whenever they do cross a branch cut, we obtain a contribution that could lead to exponential growth. (This procedure is a generalization of \cite{Roberts:2014ifa}.) We give an example in figure \ref{fig:paths}, which is for the case $x_1 < x_j < x_2$: the arrows indicate the path of the various conformal cross ratios when analytically continuing the Euclidean correlator to the OTOC. One can see that $v$, $z$ (and therefore also $\frac{u}{z}$) cross the branch cut along $(-\infty, 0]$. The contributions picked up this way give a contribution that matches the ${\cal O}(c^{-2})$ piece of the first line of \eqref{eq:answer}. We verified that other relative orderings of $x_1,x_2,x_j$ give rise to the corresponding ${\cal O}(c^{-2})$ terms of the other lines of \eqref{eq:answer}.

\bibliographystyle{utphys}
\bibliography{Operator_growth_and_black_hole_formation-v3.bbl}

\end{document}